\documentclass[aps,twocolumn,float,prd,psfig,amsmath,amssymb]{revtex4}

\usepackage[dvipdfmx]{graphicx}
\usepackage{dcolumn}
\usepackage{bm}
\usepackage{braket}
\usepackage{braket}
\usepackage{breqn}
\usepackage{color}

\begin{document}


\title{Searching anomalous polarization modes of stochastic gravitational wave background with LISA and Taiji}


\author{Hidetoshi Omiya}
\email{omiya@tap.scphys.kyoto-u.ac.jp}
\author{Naoki Seto}
\email{seto@tap.scphys.kyoto-u.ac.jp}
\affiliation{Department of Physics$,$ Kyoto University$,$ Kyoto 606-8502$,$ Japan}


\date{\today}

\begin{abstract}

The vector and scalar polarization modes of gravitational waves do not exist in General Relativity, and their detection would have  significant impacts on  fundamental physics. In this paper, we  explored the detectability of these anomalous polarization modes in a gravitational wave background around $1$ mHz with the future LISA-Taiji network.  The inherent geometrical symmetry of the network largely simplifies the correlation analysis.  By taking a suitable linear combination of the correlated outputs, the contribution of the standard tensor modes can be canceled algebraically, and  the anomalous modes can be exclusively examined. 
We provide concise expressions for the signal-to-noise ratios of the anomalous modes with this cancellation method.  We also discuss the possibility of separately estimating the amplitudes of the vector and scalar modes, using the overall  frequency dependence of the associated overlap reduction functions. 
\end{abstract}


\maketitle

\section{Introduction}
A cosmological stochastic gravitational wave background is one of the principle observational targets of  laser interferometers.  Various forms of backgrounds have been proposed to be generated in the early universe, {\it e.g.} during inflation \cite{Starobinsky:1979ty, PhysRevLett.99.221301, Cook:2011hg} and cosmological phase transitions \cite{Kamionkowski:1993fg, Caprini:2007xq,Maggiore:1999vm}. Other possible origins of backgrounds are superpositions of gravitational waves emitted by topological defects \cite{Damour:2004kw, Olmez:2010bi}, unresolved coalescing compact binaries \cite{Farmer:2003pa,PhysRevD.84.124037,Zhu:2012xw}, and so on (see \cite{Christensen:2018iqi} for a recent review). Owing to its origin, a cosmological background would be highly isotropic. 

In addition to the fact that a gravitational wave background can be used  to probe various evolutionary phases of the universe, its polarization modes  could provide an intriguing way to test  theories of gravity. General Relativity (GR) predicts gravitational waves with only two tensorial polarization modes ($+$ and $\times$ components). But, some alternative theories of gravity allow the existence of  anomalous polarization modes that are absent in GR. More precisely, we might have the following four modes; the $x$ and $y$ components for the vector modes and the $b$ and $l$ components for the scalar modes (see \cite{Will:1993ns} for their geometrical characterization). 

To detect a gravitational wave background under the presence of the detector noises, the correlation analysis is a powerful method  \cite{Flanagan:1993ix,Allen:1997ad}. By taking a cross correlation of the noise independent data streams, we can improve the statistical significance of a weak background signal. This method has been used also to detect the anomalous polarization modes (see e.g. \cite{Nishizawa:2009bf, Nishizawa:2009jh, LIGOScientific:2019vic} for laser interferometers and \cite{Cornish:2017oic} for pulsar timing array). For example, the LIGO-Virgo collaboration recently provided the upper bounds $\Omega_{GW}^V \lesssim 10^{-7}$ and $\Omega_{GW}^S \lesssim 10^{-7}$ at the frequency band $\sim 20 - 100$ Hz \cite{LIGOScientific:2019vic}. Here, $\Omega_{GW}^V$ and $\Omega_{GW}^S$ are the effective energy density spectra of the gravitational wave background for the vector and scalar modes \footnote{We provide the exact definitions of these quantities in Sec.\ref{sec:4}.}. 

The essentially new frequency band around $ 1$ mHz  will be explored by the future space-borne interferometers such as LISA \cite{Audley:2017drz}, Taiji \cite{Hu:2017mde}, and TianQin \cite{Luo:2015ght}. Each of these triangular interferometers can produce several data outputs by itself, but an intra-triangle correlation is known to be insensitive to the monopole pattern of a background due to the underlying symmetry (see e.g. \cite{Seto:2004ji}). On the other hand, we can detect the monopole pattern by taking a correlation between the different triangles. Given the rapid progress of Taiji and TianQin, it now becomes reasonable  to  assume that we can make a correlation analysis in the mHz band by using them jointly with LISA \cite{seto:xxxx}.

In this paper, we study the possibility of detecting the anomalous polarization modes in a background,  specifically with the LISA-Taiji network. As recently pointed out by Ref. \cite{seto:xxxx}, this network has a special geometrical symmetry and  the data analysis scheme of its correlation analysis can be significantly simplified.  As a result, the network provides us with just two independent correlation outputs for the even  part of the parity decomposition \cite{seto:xxxx}. Our basic strategy in this paper is to algebraically cancel the contribution of the standard tensor modes by taking an appropriate linear combination of the two outputs (see also \cite{Nishizawa:2009bf, Nishizawa:2009jh, Seto:2008sr} for related approaches). This combination is composed only of the vector and scalar modes, and confirmation of its finiteness supports  the presence of the anomalous polarization modes. For the LISA-Taiji network, in terms of  the normalized energy density spectrum, the detection limit  of the anomalous modes  will be $\sim 10^{-12}$ for a 10 yr integration. 

The outline of this paper is as follows. In section \ref{sec:2}, we describe the current orbital designs of both LISA and Taiji. Then we explain the geometry of their network and their data channels relevant for our analysis. In section \ref{sec:3}, we review the correlation analysis to detect a stochastic gravitational wave background made only with the standard tensor modes. In section \ref{sec:4}, we explain how to separate the vector and scalar polarization modes from the tensor modes. Then, we estimate the detection limit of these anomalous polarization modes with the LISA-Taiji network. We also mention the capability of simultaneous parameter estimation for the vector and scalar modes, using the Fisher matrix formalism. Our analysis up to section IV is for a fixed network configuration with a high geometrical symmetry.  In section V, we relax this  restriction. We first change the separation between two detectors, keeping the geometrical symmetry (Sec.\ref{sec:4.5A}).  Then we discuss the possibility of algebraically separating the tensor, vector, and scalar modes, by breaking the geometrical symmetry (Sec.\ref{sec:4.5B}). Finally, in section \ref{sec:5}, we summarize this paper.

\section{LISA-Taiji network}\label{sec:2}

\begin{figure}[t]
\centering
\includegraphics[keepaspectratio, scale=0.15]{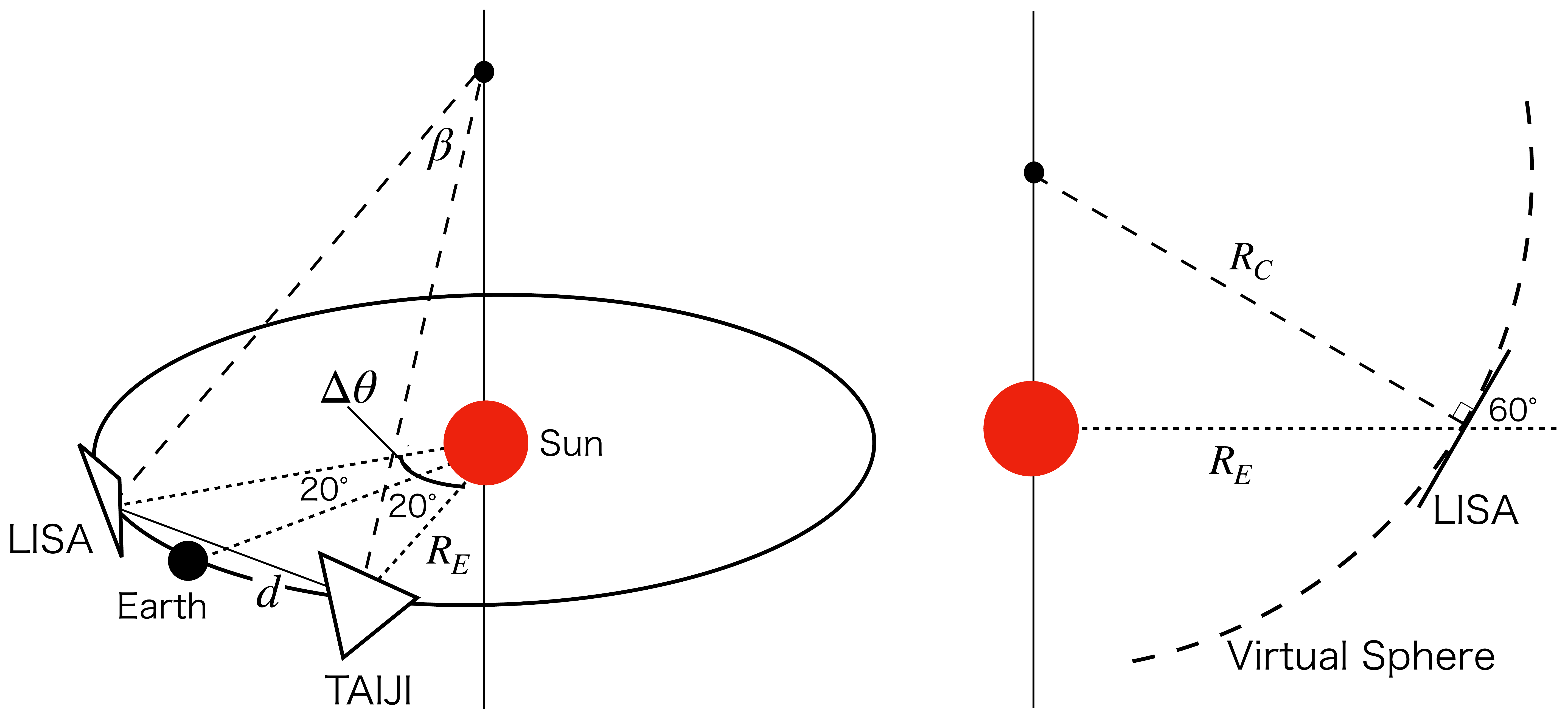}
\caption{(Left) The global geometry of the LISA-Taiji network with the orbital phase angle $\Delta \theta=40^\circ$. The virtual sphere of the radius $R_c$ is tangential to the two triangles.  Measured from the center of the virtual sphere, the opening angle between the two triangles is $\beta = 34.46^\circ$. (Right) A sectional view of the virtual sphere.  The dotted line is on the ecliptic plane with $R_E$ equal to 1AU.
}
\label{fig:1}
\end{figure}

As shown in Fig.\ref{fig:1}, LISA has a heliocentric orbit at $20^\circ$ behind the Earth. Its interferometer is composed of the three spacecraft forming a nearly equilateral triangle with the side lengths $l \sim 2.5 \times 10^6$ km. The detector plane is inclined to the orbital plane by $60^\circ$. Taiji is planned to have a similar orbital configuration ({\it e.g.} the inclination of $60^{\circ}$) as LISA. But it moves ahead of the Earth by $20^\circ$ with the arm lengths $l' \sim 3.0\times 10^6$ km. In the following, we attach  $'$  to  the quantities  related to Taiji.  

In the rest of this section, we briefly discuss the geometrical aspects of the LISA-Taiji network following \cite{seto:xxxx}. The separation between LISA and Taiji is $d = 2 R_E \sin \Delta \theta \sim 1.0 \times 10^8$ km, where $\Delta\theta=40^\circ$ is the orbital phase difference and $R_E (=1 $AU) is the mean distance from the Earth to the Sun. This separation corresponds to the frequency $c/d \sim 3$ mHz that is a key parameter for the correlation analysis with the network.  Later, in Sec.\ref{sec:4.5A}, we move the parameter $\Delta \theta$ from the planned value $40^\circ$. In this paper, we  assume that gravitational waves effectively propagate at the speed of light $c$.

\begin{figure}[t]
\centering
\includegraphics[keepaspectratio, scale=0.2]{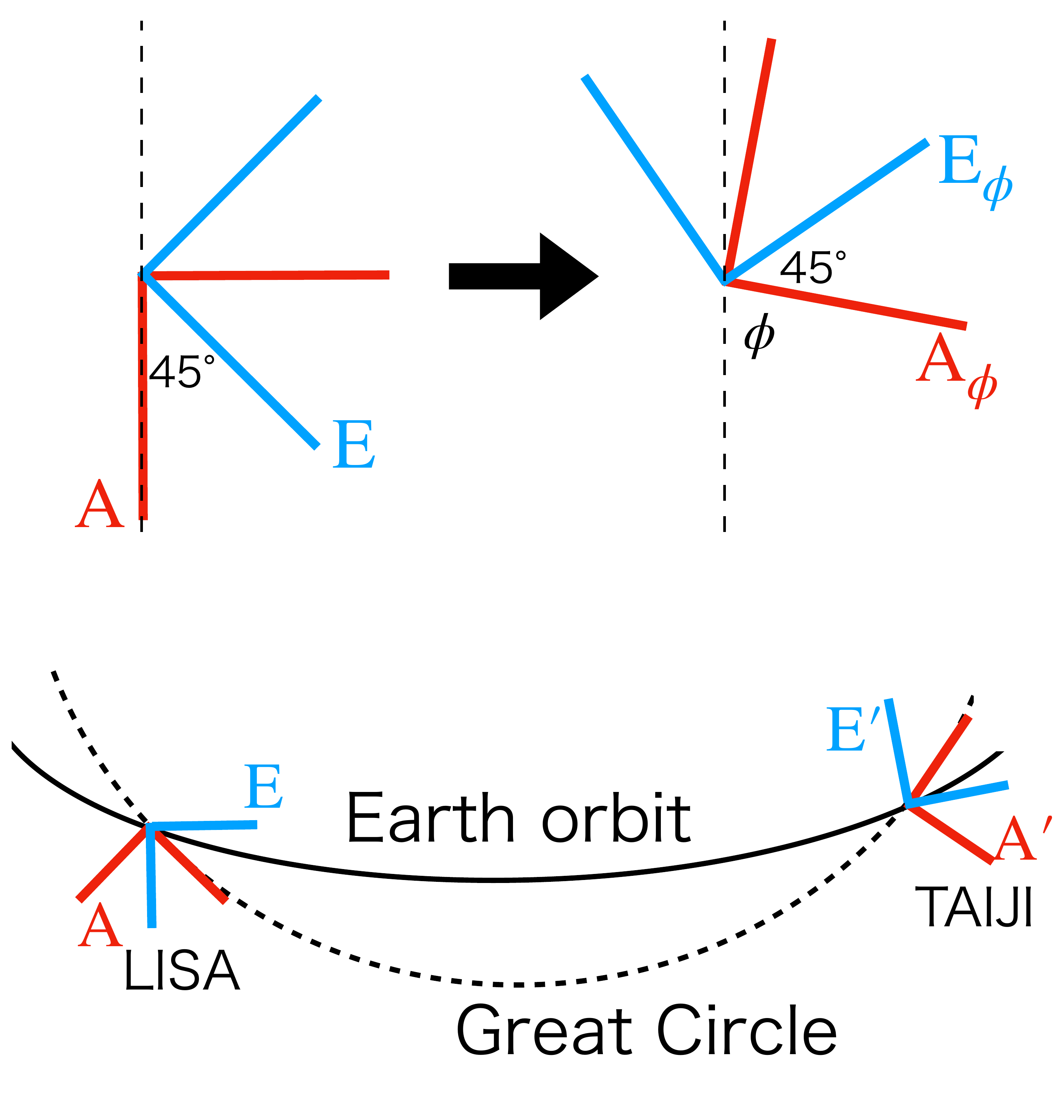}
\caption{(Top) Configuration of the two effective L-shaped interferometers A and E with the offset angle $45^\circ$ on the detector plane. By taking the data combination Eq.\eqref{eq:2new}, we can generate the new data channels $({\rm A}_\phi,{\rm E}_\phi)$ whose detector tensors are rotated by the angle $\phi$, relative to those for the original ones $\rm(A,E)$.
 (Bottom) By adjusting the rotational angle $\phi$, we arrange one arm of the A interferometer to be parallel to the great circle on the virtual sphere.}
\label{fig:1.5}
\end{figure}

Because both LISA and Taiji have the same inclination angle, their detector planes are tangent to a virtual sphere \cite{seto:xxxx} (see Fig.\ref{fig:1}). The radius of this sphere is $R_C = R_E/\sin 60^{\circ}\sim 1.15$ AU with its center above the Sun. This virtual sphere helps us easily understand the underlying symmetry of the detector network. Moreover, in relation to the correlation analysis, we can directly apply the analytic expressions originally given for the ground based-detectors that are tangent to the Earth sphere \cite{Flanagan:1993ix}. 
In Fig.\ref{fig:1}, the separation angle $\beta$ measured from the center of the sphere is given by
\begin{align}\label{eq:1}
	\beta =2 \sin^{-1}\left(\frac{d}{2 R_C}\right) \sim 34.46^\circ~.
\end{align}
In this paper, this angle will appear frequently for characterizing  the correlation between the detectors.

Next, we discuss the data channels available from the single LISA triangle.
Using the symmetry of the three vertexes, we can make the three orthogonal data channels (A, E, and T) that have independent noises \cite{PhysRevD.66.122002}. In the low frequency regime $(f\ll c/(2\pi l))$, the T channel has a negligible sensitivity compared to the A and E channels \cite{PhysRevD.66.122002}. Thus we use these two channels for our study below. Note that they have  the detector tensors equivalent to the two L-shaped interferometers with the offset angle $45^\circ$ on the detector plane (see the upper left part of Fig.\ref{fig:1.5}). We can apply the same arguments on Taiji and denote its corresponding modes by A$'$ and E$'$.

Here we should notice that the detector tensors of the A and E channels are attached to the LISA\rq{}s triangle that spins in one-year period. But, in fact,  at each epoch,  we can arbitrary rotate the two detector tensors  commonly on the detector plane,  still without noise correlation \cite{Seto:2004ji,seto:xxxx} (see the top panel of Fig.\ref{fig:1.5}). This can be attained by using the internal symmetry of the LISA\rq{}s triangle and taking the appropriate linear combinations of the original A and E channels \cite{seto:xxxx}:
\begin{align}\label{eq:2new}
	\left(
	\begin{array}{c}
	{\rm A}_\phi\\
	{\rm E}_\phi
	\end{array}\right) =
		\left(
	\begin{array}{cc}
	\cos2\phi & \sin2\phi\\
	-\sin2\phi & \cos2\phi
	\end{array}\right)
		\left(
	\begin{array}{c}
	{\rm A}\\
	{\rm E}
	\end{array}\right)~.
\end{align}
 Here, the set $({\rm A}_\phi, {\rm E}_\phi)$ is the new data channels rotated by angle $\phi$.
 
Considering the symmetry of the LISA-Taiji network elucidated by the virtual sphere, it would be reasonable to adjust the angle $\phi$ such that the new data channels $({\rm A}_\phi,{\rm E}_\phi)$ respect the great circle connecting LISA and Taiji. More specifically, for LISA, we align one arm of the interferometer ${\rm A}_\phi$ parallel to the great circle. Hereafter, for notational simplicity, we  denote the adjusted ones by $\rm (A,E)$, dropping the subscript $\phi$. We make a similar choice for Taiji (see Fig.\ref{fig:1.5}).

We have six independent data pairs, AE, A$'$E$'$, AE$'$, EA$'$, AA$'$ and EE$'$, to perform the cross correlation. But, as mentioned earlier,  the intra-triangle pairs AE and A$'$E$'$ have no sensitivity to the monopole pattern of a gravitational wave background \cite{Seto:2004ji}. 
 In addition, due to the mirror symmetry of the interferometers with respect to the plane containing the great circle, the combinations AE$'$ or EA$'$ can only probe the parity asymmetric components of an isotropic background \cite{seto:xxxx}.

So far, we have explained the basic geometrical aspects of the LISA-Taiji network, following \cite{seto:xxxx}. 
  The main topic in  that paper  was the observational decomposition  of a  tensor background  into the odd and even parity part (without considering the vector and scalar modes).   The odd parity part characterizes the asymmetry between the amplitudes of the right- and left-handed circularity polarized waves. In contrast,   the even part shows the summation of the two amplitudes, or equivalently the total intensity. Our main topic in this paper is  the detectability of the vector and scalar  polarization modes with no parity asymmetry. Therefore, except for Sec.\ref{sec:4.5B} where the mirror symmetry is no longer applicable, we can focus our study on the  even parity pairs AA$'$ and EE$'$.

\section{Correlation Analysis}\label{sec:3}

The correlation analysis is a powerful method to detect a stochastic gravitational wave background \cite{Flanagan:1993ix,Allen:1997ad}. Here, we review this method, targeting a gravitational background purely made with the parity-symmetric tensor modes (assuming GR). We derive basic expressions that will be used in the next section for the anomalous polarization search.

First, we decompose the metric perturbation induced by a stationary, isotropic and independently polarized gravitational wave background
 as
\begin{align}\label{eq:2}
\begin{aligned}
	h_{ij}(t,\bm{x}) = &\sum_{P= +,\times} \int df \int d\bm{\Omega}\\
	& \times \tilde{h}_P(f,\bm{\Omega}) \bm{e}_{P,ij}(\bm{\Omega}) e^{2\pi i f (t - \bm{\Omega} \cdot \bm{x}/c)}~.
\end{aligned}
\end{align}
Here, the unit vector $\bm{\Omega}$ is defined on the two sphere, and the polarization tensor $e_{P}$ takes the $+$ and $\times$ components for GR. We defined the solid angle element $d\bm{\Omega}$, such that $\int d\bm{\Omega} = 4 \pi$ for the surface integral on a unit sphere. 
The explicit form of the tensors  $e_{+}$ and $e_{\times}$ are given by
\begin{align}\label{eq:3}
\begin{aligned}
	\bm{e}_{+}(\bm{\Omega}) &= \bm{m} \otimes \bm{m} - \bm{n}\otimes \bm{n}~\\
	 \bm{e}_{\times}(\bm{\Omega}) &= \bm{m} \otimes \bm{n} + \bm{n}\otimes \bm{m}~,
\end{aligned}
\end{align}
where $(\bm{m}, \bm{n}, \bm{\Omega})$ forms an orthonormal basis (see \cite{Nishizawa:2009jh} for their detail).

In Eq.\eqref{eq:2}, $\tilde{h}_P$ are the mode coefficients and their statistical properties are determined by the power spectrum density as
\begin{align}\label{eq:4}
	\braket{\tilde{h}_{P}(f,\bm{\Omega})\tilde{h}_{P'}^*(f',\bm{\Omega'})} &= \delta_{PP'} \delta_{\Omega\Omega'}\delta(f-f')S_h^T(f)~
\end{align}
with $P,P' = +,\times$. The delta function $\delta(f-f')$ follows from the stationarity of the  background. We will omit this factor for notational simplicity, but recover it if needed. 
The power spectrum density $S_h^T$ is written by $\Omega_{GW}$, which is the energy density of the gravitational waves per unit logarithmic frequency and  is normalized by the critical density of the universe \cite{Allen:1997ad}. In GR, we only have the tensor modes with the relation
\begin{align}\label{eq:5}
	\Omega_{GW}^T(f) = \left(\frac{32\pi^3}{3 H_0^2}\right) f^3 S^T_h(f)~.
\end{align}
Here, $H_0$ is the Hubble parameter and we use $H_0 = 70$ km\ s$^{-1}$\ Mpc$^{-1}$ in this paper. Note this relation might be changed for alternative theories of gravity \cite{Isi:2018miq}.

Now we discuss the relevant data channels for LISA (A and E) and Taiji (A$'$ and E$'$) in Fourier space. Each of them $s_a(f)$ ($a =$ A, E, A$'$, and E$'$) is assumed to be the sum of the background signal $h_a(f)$ and the instrumental noise $n_a(f)$:
\begin{align}\label{eq:6}
	s_a(f) = h_a(f) + n_a(f)~.
\end{align}
If the wavelength of a gravitational wave is much larger then the arm length of the interferometer, $h_a$ is simply modeled by
\begin{align}\label{eq:7}
	h_a(f) = \bm{D}_{a}^{ij} \tilde{h}_{ij}(f, \bm{x}_a)~.
\end{align}
Here $\bm{x}_a$ is the position of the interferometer, $\tilde{h}_{ij}(f)$ is Fourier transformation of $h_{ij}(t)$, and $\bm{D}_a$ is the detector tensor which represents the response of the interferometer to the incident gravitational wave \cite{Flanagan:1993ix}. The arm length of LISA and Taiji is around $l \sim l' \sim 3\times 10^6$ km, and therefore the low frequency approximation is valid at $f \lesssim  c/(2\pi l) \sim 0.02$ Hz.

In terms of the unit vectors $\bm{u}$ and $\bm{v}$ for the arm directions of the A interferometer, $\bm{D}_{A}$ is given by 
\begin{align}
	\bm{D}_{A} = \frac{1}{2}\left(\bm{u}\otimes \bm{u} - \bm{v}\otimes\bm{v} \right)~.
\end{align}
Using the same vectors, we have
\begin{align}
	\bm{D}_{E} = \frac{1}{2}\left(\bm{u}\otimes \bm{v} + \bm{v}\otimes\bm{u} \right)~
\end{align}
for the $E$ channel \cite{seto:xxxx}. We can make a similar decompositions $\bm{D}_{\mathrm{A}'}$ and $\bm{D}_{\mathrm{E}'}$ for Taiji.

The statistical properties of the instrumental noise is characterized by the noise spectrum $N_a(f)$. After dropping the delta function $\delta(f-f')$ as mentioned after Eq.\eqref{eq:4}, we obtain
\begin{align}
	\braket{n_a(f)n_b^*(f)} = \frac{1}{2}\delta_{ab} N_a(f)~.
\end{align}
Owing to the symmetry of the network, the four data streams are assumed to have independent noises, and we can put $N_A(f) = N_E(f) = N(f)$ for LISA and $ N_{A'}(f) = N_{E'}(f) = N'(f)$ for Taiji (for their analytic expressions see Ref.\cite{Cornish:2018dyw} for LISA and Ref.\cite{Wang:2020vkg} for Taiji).

As we discussed in Sec.\ref{sec:2} for the LISA-Taiji network, we only have two data pairs, AA$'$ and EE$'$ that are non-vanishing for the even parity  part. We define the expectation value for the cross correlation of the  two data pairs
\begin{align}\label{eq:11}
	C_{ab}(f) \equiv \braket{s_a(f) s_b^*(f)} = \braket{h_a(f) h_b^*(f)}
\end{align}
with $(a,b) = (\mathrm{AA}')$ or $(\mathrm{EE}')$. We used independence of the instrumental noises $\braket{n_a(f)n_b^*(f)} = 0$ in the last equality of Eq.\eqref{eq:11}. Using Eqs.\eqref{eq:4}, \eqref{eq:6}, and \eqref{eq:11}, we obtain
\begin{align}\label{eq:12}
	C_{ab}(f) = C_{ab}^T(f) \equiv \frac{8\pi}{5}\gamma^T_{ab}(f) S_h^T(f)~.
\end{align}
Here $C_{ab}^T(f)$ is the expectation value only by the tensor modes. We also introduced the overlap reduction function
\begin{align}\label{eq:13}
\begin{aligned}
	\gamma^{T}_{ab}&(f) \equiv \\
	& \frac{5}{8 \pi}\sum_{P= +,\times} \int d\bm{\Omega} \ \bm{D}_{a,ij}\bm{D}_{b,kl}\bm{e}^{ij}_{P}\bm{e}^{kl}_{P} e^{2 \pi i f \bm{\Omega}\cdot (\bm{x}_{a}-\bm{x}_b)/c}~
\end{aligned}
\end{align}
for a background purely made with the tensor modes. 
It quantifies the correlated responses of the detectors to the background signal \cite{Flanagan:1993ix, Allen:1997ad}.

\begin{figure}[h]
\centering
\includegraphics[keepaspectratio, scale=0.6]{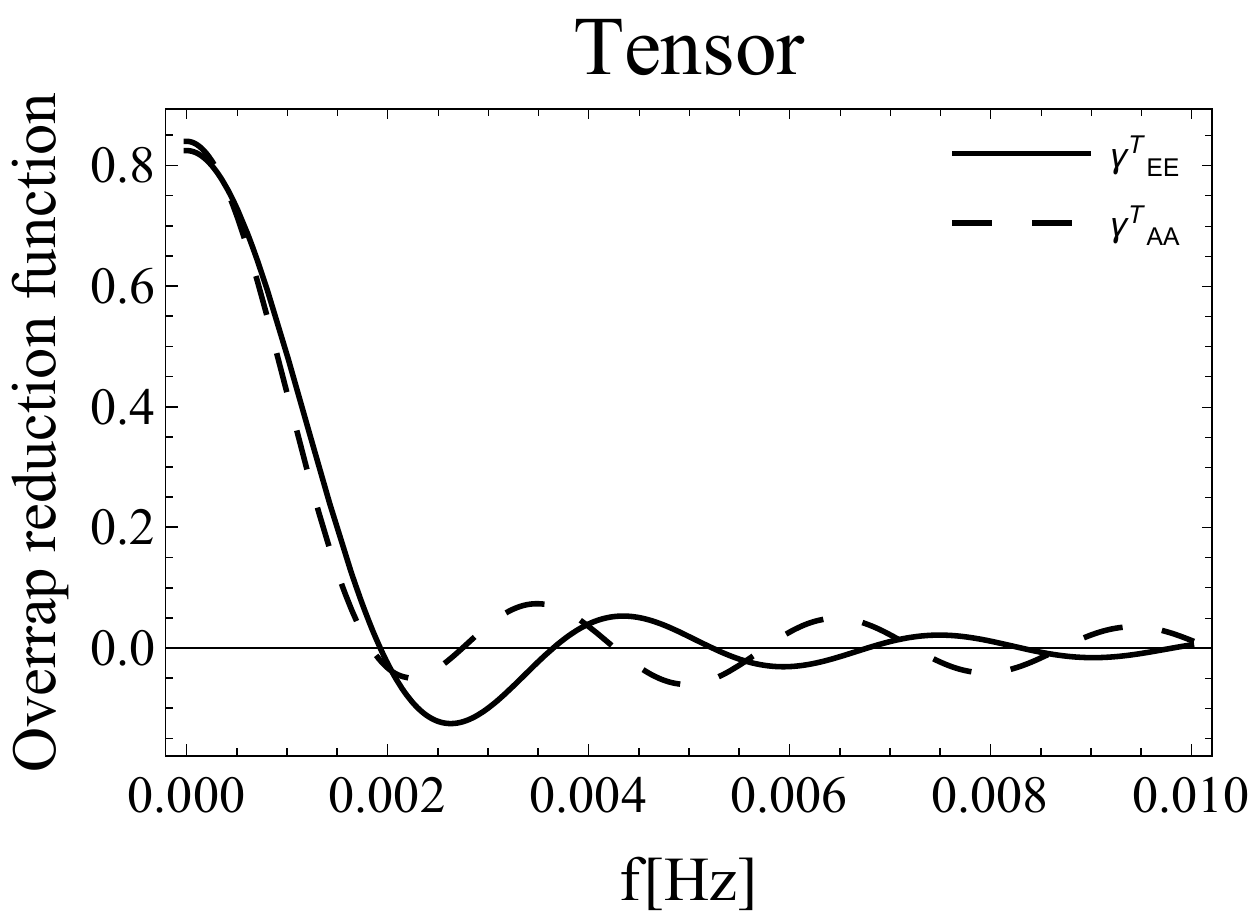}
\caption{The overlap reduction functions of the tensor modes for the LISA-Taiji network. The solid and dashed lines correspond to the $\mathrm{EE'}$ and the $\mathrm{AA'}$ data pairs, respectively.}
\label{fig:3}
\end{figure}

Using the literature for the ground-based networks \cite{Flanagan:1993ix}, we obtain
\begin{align}
\label{eq:14}
	\gamma^T_{AA'} &= \Theta_1^T(y,\beta) - \Theta_2^T(y,\beta)~,\\
	\gamma^T_{EE'} &= \Theta_1^T(y,\beta) + \Theta_2^T(y,\beta)~,
\end{align} 
with
\begin{gather}
	\Theta_1^T(y,\beta) = \left(j_{0}(y) + \frac{5}{7}j_2(y) + \frac{3}{112} j_4(y)\right)\cos^4\left(\frac{\beta}{2}\right)\\
	\label{eq:17}
	\begin{aligned}
	\Theta_2^T(y,\beta) &= \left(-\frac{3}{8}j_0(y) + \frac{45}{56}j_2(y) - \frac{169}{896} j_4(y)\right)\\
	&+\left(\frac{1}{2}j_0(y) - \frac{5}{7}j_2(y) - \frac{27}{224} j_4(y)\right)\cos\beta\\
	&+\left(-\frac{1}{8}j_0(y) - \frac{5}{56}j_2(y) - \frac{3}{896} j_4(y)\right)\cos2\beta~.
	\end{aligned}
\end{gather}
Here, $j_{n}$ are the spherical Bessel functions with their arguments $y = 2 \pi  fd/c$. For the LISA-Taiji network, the opening angle $\beta$ is $34.46^\circ$ and distance between the triangles is $d \sim 1.0\times10^8$ km  (see Fig.\ref{fig:1}). In Fig.\ref{fig:3}, we show the two overlap reduction functions in the low frequency regime.

We briefly discuss the asymptotic behaviors of the overlap reduction functions at the small and large frequency regimes. Using the property of the spherical Bessel function
\begin{align}
	j_{l}(x) &\underset{x\to 0}{\to} \frac{2^l l!}{(2l + 1)!} x^l~,
\end{align}
we can show
\begin{align}\label{eq:19}
\begin{aligned}
	\lim_{f\to 0}\gamma_{ab}^T &= D_{a,ij}D_{b}^{ij}/2~,
\end{aligned}
\end{align}
which is unity when two detectors are coincident and aligned (namely $a=b$) \cite{Flanagan:1993ix}. For the LISA and Taiji network, we obtain
\begin{align}\label{eq:20}
\begin{aligned}
	\lim_{f\to 0}\gamma_{AA'}^T = \cos^4(\beta/2) + \sin^4(\beta/2) = 0.840~,\\
	\lim_{f\to 0}\gamma_{EE'}^T = \cos^4(\beta/2) - \sin^4(\beta/2) = 0.825~.
\end{aligned}
\end{align}

In the large frequency regime, the spherical Bessel functions behave as 
\begin{align}
	j_{l}(x) &\underset{x\to \infty}{\to} \frac{1}{x} \cos(x - (l+1)\frac{\pi}{2})~.
\end{align}
Thus in Fig.\ref{fig:3} the overlap reduction functions oscillate with the frequency interval $c/d \sim 3 \mathrm{mHz}$ at $f \gtrsim 5$ mHz.

In Fig.\ref{fig:3}, we simultaneously have $\gamma^T_{AA'} \sim \gamma^T_{EE'}\sim 0$ around 2 mHz. This is just a coincidence realized at the specific angle $\beta =  34.46^\circ$, and it causes some interesting results in section \ref{sec:4}.

\section{Anomalous Polarization search}\label{sec:4}

In the previous section, we only considered a background purely made with the tensor modes.  But,  in the alternative theories of gravity, a background  could also contain the vector and scalar modes. In this section, we investigate the contribution of these anomalous modes and discuss how to detect them separately from the standard tensor modes, using the LISA-Taiji network. In Sec.\ref{sec:4A}, we explain our basic idea for the anomalous mode search after eliminating the tensor modes. Then, we discuss a background composed of the tensor and vector modes (Sec.\ref{sec:4B}), and the tensor and scalar modes (Sec.\ref{sec:4C}). In section \ref{sec:4D} we examine a background simultaneously made with the three polarization modes, and discuss the decomposition of the vector and the scalar modes using the frequency dependence of the overlap reduction functions.

\subsection{Elimination of the tensor modes}\label{sec:4A}

Let us consider the following data combination for the LISA-Taiji network:
\begin{align}\label{eq:22}
	\mu \equiv \gamma_{EE'}^T s_{A}(f)s^*_{A'}(f) - \gamma_{AA'}^T s_{E}(f) s_{E'}^*(f)~.
\end{align}
Here, $ \gamma_{EE'}^T $ and $\gamma_{AA'}^T$ should be regarded as the known coefficients calculated theoretically.
Using Eqs.\eqref{eq:6}, and \eqref{eq:11}, we obtain the expectation value
\begin{align}
\begin{aligned}\label{eq:23}
	\braket{\mu} &= \gamma^T_{EE'} \braket{h_{A} h^{*}_{A'}} - \gamma^T_{AA'} \braket{h_{E} h^*_{E'}}\\
	&= \gamma^T_{EE'}(f)C_{AA'}(f) - \gamma^T_{AA'}(f)C_{EE'}(f)~.
\end{aligned}
\end{align}
In the first equality, we used independence of the instrumental noises. If the background is purely made with the tensor modes, we have 
\begin{align}\label{eq:24}
	C_{ab} = C_{ab}^T = \frac{8\pi}{5}\gamma^T_{ab}(f) S_h^T(f)~
\end{align}
as in Eq.\eqref{eq:12}. Substituting Eq.\eqref{eq:24} into Eq.\eqref{eq:23}, we obtain
\begin{align}
	\braket{\mu}\bigl|_{T} = 0~.
\end{align}
Here, $\braket{\cdot}\bigl|_{T}$ represents the expectation value for a background only with the tensor modes. However, under the presence of the additional polarization modes, we obtain $\braket{\mu}\neq 0$, still algebraically eliminating the contribution of the tensor modes. We will calculate the expectation value $\braket{\mu}$ after evaluating the overlap reduction functions for the vector and scalar modes.

At this point, let us calculate the statistical fluctuations for the data combination $\mu$. Here, following the standard arguments on the correlation analysis, we assume that the background signal is much smaller than the instrumental noise $|h_a| \ll |n_a|$. Then for the data combination $\mu$, the variance $\sigma_\mu^2$ is given by
\begin{align}\label{eq:26}
\begin{aligned}
	\sigma_{\mu}(f)^2 &\sim \frac{1}{4}\left(\left(\gamma^{T}_{EE'}(f)\right)^2 + \left(\gamma^{T}_{AA'}(f)\right)^2 \right) N(f) N'(f)
\end{aligned}
\end{align}
 (see \cite{Seto:2005qy} for detail of the derivation). Recalling our prescription for the delta function and summing up all the frequency segments, we obtain the signal-to-noise ratio
\begin{align}\label{eq:27}
	\mathrm{SNR}^2 = \int_{f_{cut}}^{\infty} df \frac{\braket{\mu}^2}{\sigma_{\mu}^2}~,
\end{align}
as in \cite{Seto:2005qy}. Here, we introduced the low-frequency cut off $f_{cut}$ to take into account the potential contamination of the Galactic binary confusion noise \cite{Seto:2005qy}. The actual value of the $f_{cut}$ would depend on the mission lifetimes of LISA and Taiji.

\subsection{Vector modes}\label{sec:4B}

Next we consider a background made of the tensor and vector modes (without the scalar modes), and discuss the isolation of the later. The vector modes are characterized by the following polarization tensors:
\begin{align}\label{eq:28}
\begin{aligned}
	\bm{e}_{x} &= \bm{\Omega} \otimes \bm{m} + \bm{m}\otimes \bm{\Omega}~,& \bm{e}_y &= \bm{\Omega} \otimes \bm{n} + \bm{n}\otimes \bm{\Omega}~,
\end{aligned}
\end{align}
where the unit vectors $\bm{\Omega}, \bm{m}$ and $\bm{n}$ are the same as those in Eq.\eqref{eq:3}. Hereafter, we assume that the vector components are independently polarized. 

As in the case of the tensor components, the statistical properties of the vector background are characterized by the power spectrum density given by
\begin{align}\label{eq:29}
	\braket{h_{P}(f,\bm{\Omega})h_{P'}^*(f,\bm{\Omega'})} &= \delta_{PP'} \delta_{\Omega\Omega'}S_h^V(f)
\end{align}
with the index $P$ and $P'$ for the two polarization states $x$ and $y$. Following Eq.\eqref{eq:5}, we introduce the effective energy density $\tilde{\Omega}^V_{GW}(f)$ by
\begin{align}\label{eq:30}
	\tilde{\Omega}_{GW}^{V}(f) \equiv \left(\frac{32\pi^3}{3 H_0^2}\right) f^3 S^{V}_{h}(f)~
\end{align}
to parametrize the strength of the vector background. We should notice that the quantity $\tilde{\Omega}^V_{GW}$ does not always represent the actual energy density $\Omega_{GW}$. The relation between the strain spectrum $S_h^V(f)$ and the energy density depends on the details of the gravitational theories under consideration \cite{Isi:2018miq}  (see Appendix).

\begin{figure}
\centering
\includegraphics[keepaspectratio, scale=0.6]{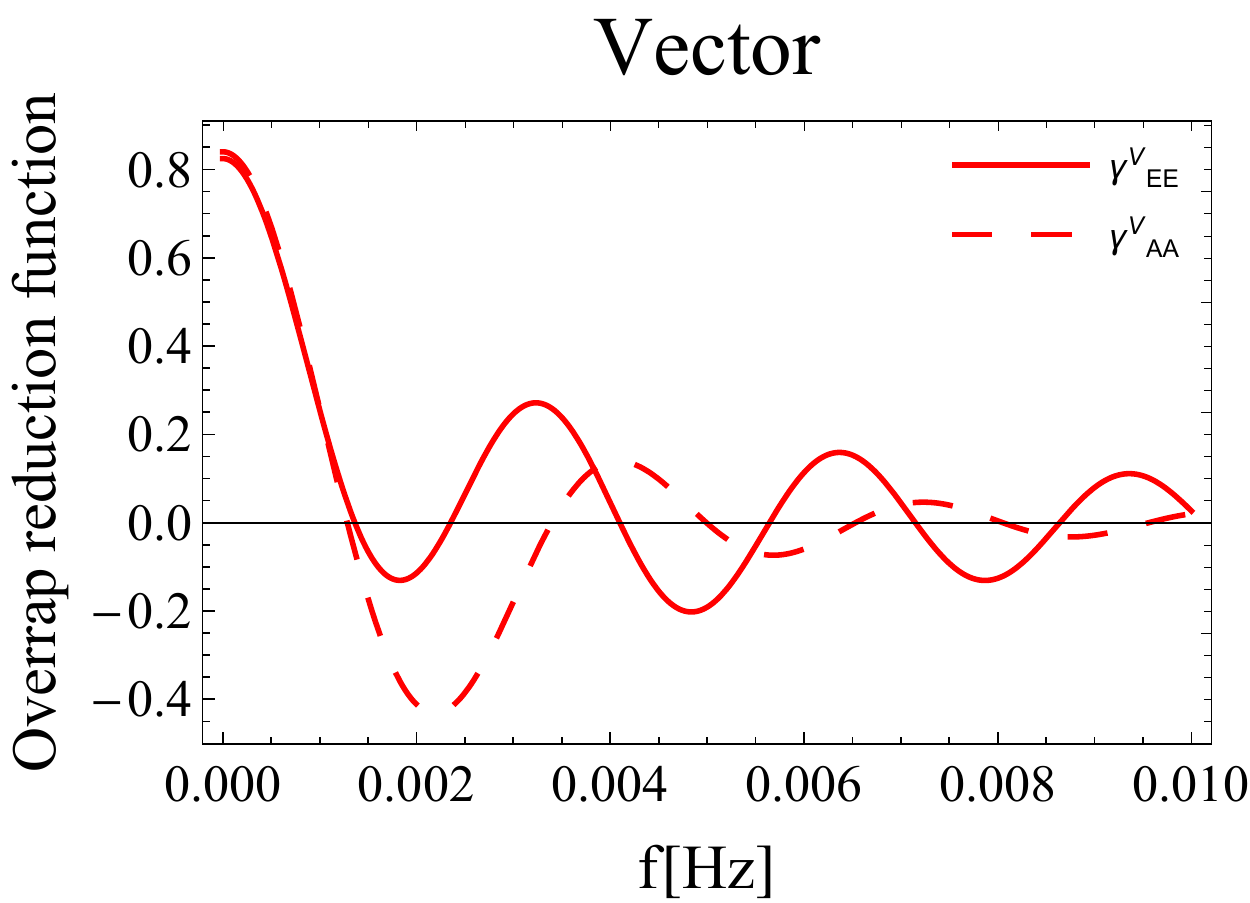}
\caption{The overlap reduction functions of the LISA-Taiji network for the vector modes. The solid and dashed lines correspond to the $\mathrm{EE'}$ and $\mathrm{AA'}$ data pairs, respectively.}
\label{fig:4}
\end{figure}

Now we calculate the cross correlation of the two data channels in the same way as in Eq.\eqref{eq:12}. For the tensor and vector blended background, we have
\begin{align}\label{eq:31}
\begin{aligned}
	C_{ab}(f) &= C_{ab}^{TV}(f) \\
	&\equiv \frac{8\pi}{5}\left(\gamma^T_{ab}(f) S_{h}^T(f) + \gamma^V_{ab}(f) S_{h}^V(f)\right)~,
\end{aligned}
\end{align}
where $\gamma^V_{ab}$ is the overlap reduction function for the vector modes. It can be evaluated by the replacement $(+,\times)\to (x,y)$ in Eq.\eqref{eq:13}. As in Eqs.\eqref{eq:14} - \eqref{eq:17} for the tensor modes, the functions $\gamma^V_{AA'}$ and $\gamma^V_{EE'}$ are written by the spherical Bessel functions as the followings \cite{Nishizawa:2009bf}:
\begin{align}\label{eq:32}
	\gamma^V_{AA'} &= \Theta_1^V(y,\beta)- \Theta_2^V(y,\beta)~\\
	\gamma^V_{EE'} &= \Theta_1^V(y,\beta)+ \Theta_2^V(y,\beta)~
\end{align} 
with
\begin{gather}
	\Theta_1^V(y,\beta) = \left(j_{0}(y) - \frac{5}{14}j_2(y) - \frac{3}{28} j_4(y)\right)\cos^4\left(\frac{\beta}{2}\right)\\
	\label{eq:35}
	\begin{aligned}
	\Theta_2^V(y,\beta) &= \left(-\frac{3}{8}j_0(y) + \frac{45}{112}j_2(y) - \frac{169}{224} j_4(y)\right)\\
	&+\left(\frac{1}{2}j_0(y) + \frac{5}{14}j_2(y) + \frac{27}{56} j_4(y)\right)\cos\beta\\
	&+\left(-\frac{1}{8}j_0(y) + \frac{5}{112}j_2(y) + \frac{3}{224} j_4(y)\right)\cos2\beta~.
	\end{aligned}
\end{gather}
In Fig.\ref{fig:4}, we show the overlap reduction functions of the vector modes for the $\mathrm{AA'}$ and $\mathrm{EE'}$ data pairs. 

In the low frequency limit $f \to 0$, we have $\gamma^V_{ab} = D_{a,ij}D_{b}^{ij}/2$ that is identical to the tensor modes $\gamma^T_{ab}$, as shown in Eqs.\eqref{eq:19} and \eqref{eq:20}. Also, their high-frequency behaviors are qualitatively similar to the tensor modes. At $f \gtrsim 5$ mHz we can again observe wavy profiles with the frequency interval $c/d \sim 3$mHz.

After substituting Eq.\eqref{eq:31} into Eq.\eqref{eq:23}, for the blended background, the expectation value of our estimator $\mu$ is given by
\begin{align}\label{eq:36}
\begin{aligned}
	\braket{\mu}\bigl|_{T,V} &= \frac{8\pi}{5}\left[\gamma^T_{EE'}(f)\gamma^V_{AA'}(f) - \gamma^T_{AA'}(f)\gamma^V_{EE'}(f)\right]S_{h}^V(f)~.
\end{aligned}
\end{align}
In general, the bracket $[\cdots]$ is non-vanishing, and we can isolate the vector modes by canceling the tensor modes.

\begin{figure}
\centering
\includegraphics[keepaspectratio, scale=0.6]{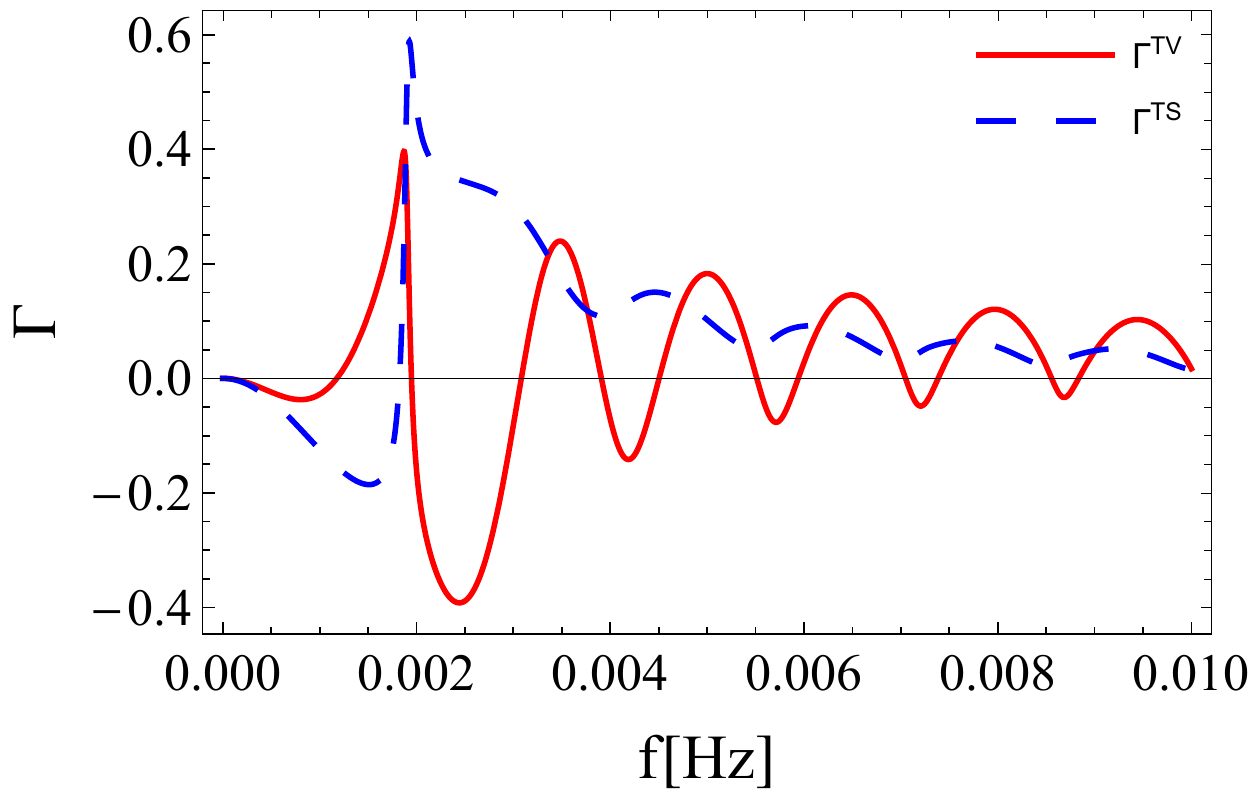}
\caption{The effective overlap reduction functions for the vector and the scalar modes given in Eqs.\eqref{eq:38} and \eqref{eq:52}. The red solid and blue dashed curves correspond to the vector and the scalar compiled overlap reduction functions, respectively. }
\label{fig:5}
\end{figure}

Next we evaluate the signal-to-noise ratio of the vector modes with our estimator $\mu$. Using Eqs.\eqref{eq:26}, \eqref{eq:27}, \eqref{eq:30}, and \eqref{eq:36}, the signal-to-noise ratio is formally given by
\begin{align}\label{eq:37}
\begin{aligned}
	\mathrm{SNR}_{V}^2(f_{cut}) =& \left(\frac{3 H_0^2}{10\pi^2}\right)^2 T_{obs} \\
	&\times \left[2\int_{f_{cut}}^{\infty} df \frac{\left(\Gamma^{TV}(f)\tilde{\Omega}_{GW}^V(f)\right)^2}{f^6N(f)N'(f)}\right]~,
\end{aligned}
\end{align}
with the effective overlap reduction function defined by
\begin{equation}\label{eq:38}
	\Gamma^{TV}(f) \equiv \frac{\gamma^T_{EE'}(f)\gamma^V_{AA'}(f) - \gamma^T_{AA'}(f)\gamma^V_{EE'}(f)}{\sqrt{\left(\gamma^{T}_{AA'}(f)\right)^2 + \left(\gamma^{T}_{EE'}(f)\right)^2}}~.
\end{equation}
In Fig.\ref{fig:5} we present $\Gamma^{TV}(f)$ in the frequency regime appropriate for the low frequency approximation. We see the sudden change of $\Gamma^{TV}$ around 2 mHz. This is due to the proximity of the zero points of the two functions $\gamma^T_{AA'}$ and $\gamma^T_{EE'}$, as shown in Fig.\ref{fig:3}. In Fig.\ref{fig:5}, the function $\Gamma^{TV}(f)$ rapidly decays below $ f = 2$ mHz, reflecting the property $\gamma^T_{ab}(y) \sim \gamma^V_{ab}(y)$ around $y = 0$. At $f \gtrsim 2$ mHz, we can also observe the oscillation with the interval $c/2d \sim 1.5 \mathrm{mHz}$. 

The formal expression Eq.\eqref{eq:37} is  given as the optimal signal-to-noise ratio. It can be evaluated, once we assume the actual model for the spectrum $\tilde{\Omega}_{GW}^V$. Below, for simplicity, we suppose that the true vector background has a flat spectrum $\tilde{\Omega}_{GW}^V(f) = \tilde{\Omega}_{GW}^V$. After numerically integrating Eq.\eqref{eq:37}, we can express the result in the following form:
\begin{align}\label{eq:39}
	\mathrm{SNR}_{V}(f_{cut}) &= 17.3\left(\frac{\tilde{\Omega}_{GW}^V}{10^{-12}}\right) \left(\frac{T_{obs}}{10 \mathrm{yr}}\right)^{1/2} \mathcal{F}_V(f_{cut})~.
\end{align}
Here $\mathcal{F}_V(f_{cut})$ shows the dependence on the cut-off frequency $f_{cut}$ with the normalization
\begin{align}
	\mathcal{F}_V(0) = 1~.
\end{align}
We evaluated our numerical results, assuming a 10 yr observation, i.e. $T_{obs} = 10$ yr, which is the maximum operation time argued for LISA. This would be a highly optimistic choice for the LISA-Taiji network, but we can easily scale our results for different values of $T_{obs}$. For correlation analysis, we can use only the perfectly overlapped period of two detectors. To ensure a large integration time $T_{obs}$, a coordinated operation schedule (e.g. maintenance time, etc) would be advantageous.

\begin{figure}
\centering
\includegraphics[keepaspectratio, scale=0.6]{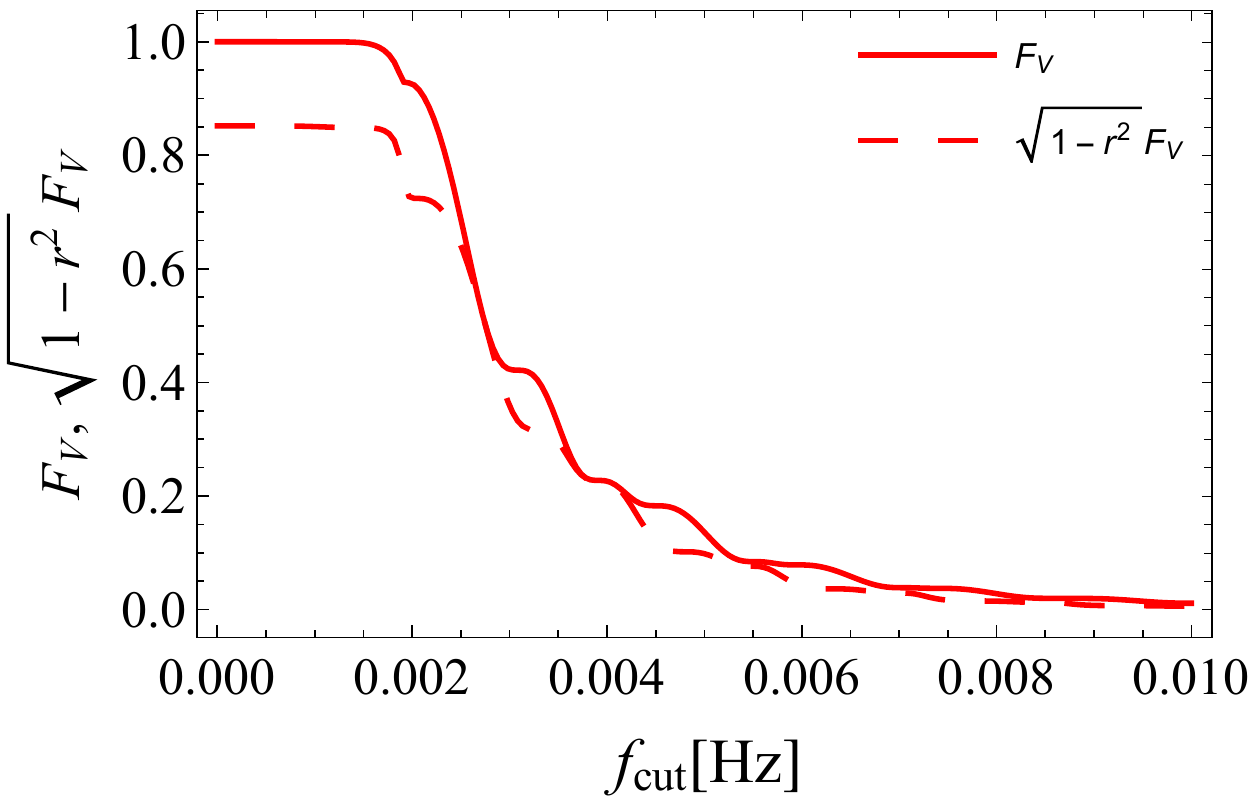}
\caption{Dependence of statistical quantities for the vector modes on the frequency cutoff $f_{cut}$. The solid line represents the function $\mathcal{F}_V(f_{cut})$ for the signal-to-noise ratio in Eq.\eqref{eq:39}, and for the estimation error  in Eq.\eqref{eq:41}. The dashed line is for the two dimensional parameter estimation in Eq.\eqref{eq:66} with $P = V$. The factor $\sqrt{1-r^2}$ shows the statistical loss by the covariance of two parameters.}
\label{fig:6}
\end{figure}

In Fig.\ref{fig:6}, we show the function $\mathcal{F}_V(f_{cut})$. The step-like profile above 2 mHz is caused by the oscillation of $\Gamma^{TV}(f)$ shown in Fig.\ref{fig:5}. We can also find that the signal-to-noise ratio is less sensitive to $f_{cut}$ below 2 mHz, mainly due to the suppression of $\Gamma^{TV}(f)$ there. Fig.\ref{fig:5} indicates that for $\mathrm{SNR}_V$, the contribution of $f \gtrsim c/(2\pi l) \sim 0.02$ Hz is totally negligible. This justifies our evaluations based on the low frequency approximation.

Now let us consider a situation that we estimate the amplitude $\tilde{\Omega}_{GW}^V$ of the flat spectrum by applying the standard maximum likelihood analysis to our estimator $\mu$. Using the Fisher matrix approach to the single fitting parameter $\tilde{\Omega}_{GW}^V$, we obtain the relative error \cite{Seto:2005qy}
\begin{align}\label{eq:41}
	\Braket{\left(\frac{\Delta \tilde{\Omega}_{GW}^V}{\tilde{\Omega}_{GW}^V}\right)^2}^{1/2} &= \frac{1}{\mathrm{SNR}_V(f_{cut})}\\
	\label{eq:42}
	& \propto \frac{1}{\mathcal{F}_V(f_{cut})}
\end{align}
(see also Ref.\cite{Allen:1997ad}). Later in Sec.\ref{sec:4D}, we deal with a more complicated case for  simultaneously estimating the multiple parameters.

\subsection{Scalar modes}\label{sec:4C}

Next we consider a background made with the tensor and the scalar modes but without the vector modes. The polarizations of the scalar modes are characterized by the following two tensors:
\begin{align}\label{eq:43}
\begin{aligned}
	\bm{e}_b &= \sqrt{3}(\bm{m} \otimes \bm{m} + \bm{n}\otimes \bm{n})~,& \bm{e}_l &= \sqrt{3}(\bm{\Omega}\otimes\bm{\Omega})~.
\end{aligned}
\end{align}
The subscripts $b$ and $l$ denote the breathing and the longitudinal modes, respectively (see Appendix  for the explanation of the unconventional factor of $\sqrt{3}$ ). 
As in Eqs.\eqref{eq:4} and \eqref{eq:29}, we introduce the power spectrum density by
\begin{align}
	\braket{h_{P}(f,\bm{\Omega})h_{P'}^*(f,\bm{\Omega'})} &= \delta_{PP'} \delta_{\Omega\Omega'}S_h^P(f)~.
\end{align}
Here, the indexes $P$ and $P'$ denote the two polarization states ($b$ and $l$) that are assumed to be statistically independent.

In a similar way as the vector modes, we define the effective energy density $\tilde{\Omega}_{GW}^S$ of the scalar background by
\begin{align}\label{eq:45}
	\tilde{\Omega}_{GW}^{S}(f) \equiv \left(\frac{32\pi^3}{3 H_0^2}\right) f^3 S^{S}_{h}(f)~,
\end{align}
where $S^S_h(f) \equiv (S_h^b(f) + S_h^l(f))/2$ is the mean power spectrum of the scalar modes. Also for the scalar modes, the effective energy density $\tilde{\Omega}_{GW}$ could be different from the actual energy density (see Sec.\ref{sec:4B} for the discussion on the vector modes).

Now we calculate the expectation value of our estimator $\mu$ for the background composed of the tensor and scalar modes. Following the same steps to derive Eq.\eqref{eq:36} for the tensor-vector blended background, we obtain
\begin{align}\label{eq:46}
\begin{aligned}
	\braket{\mu}\bigl|_{T,S} &= \frac{8\pi}{5}\left[\gamma^T_{AA'}(f)\gamma^S_{EE'}(f) - \gamma^T_{EE'}(f)\gamma^S_{AA'}(f)\right]S_{h}^S(f)~.
\end{aligned}
\end{align}
Here, $\gamma^S_{ab}$ are the overlap reduction functions for the scalar modes. As in the case of the tensor and vector modes (see Eq.(\ref{eq:13})),  we defined them as the summation of the contributions from the breathing and longitudinal modes. But actually, they have  identical overlap reduction functions. This can be understood as follows. From Eq.\eqref{eq:43}, the summations $\bm{e}_b + \bm{e}_l$ is proportional to the unit matrix. In addition, the detector tensor $D_{a}^{ij}$ is traceless and we obtain the resultant relation $D_{a}^{ij} e_{b,ij} =-D_{a}^{ij} e_{l,ij}$. Applying this relation to the integrals corresponding to Eq.\eqref{eq:13}, the overlap reduction functions for the breathing and longitudinal modes become the same \cite{Chatziioannou:2012rf,Nishizawa:2009bf}.   Accordingly, only the mean spectrum $S_{h}^S$ appears in Eq.\eqref{eq:46}.

 The explicit expressions for the overlap reductions functions  are obtained by using expressions in  \cite{{Nishizawa:2009bf}} as follows
\begin{align}
	\gamma^S_{AA'} &= \Theta_1^S(y,\beta)- \Theta_2^S(y,\beta)\\
	\gamma^S_{EE'} &= \Theta_1^S(y,\beta)+ \Theta_2^S(y,\beta)
\end{align} 
with
\begin{gather}
	\Theta_1^S(y,\beta) = \left(j_{0}(y) - \frac{5}{7}j_2(y) + \frac{9}{56} j_4(y)\right)\cos^4\left(\frac{\beta}{2}\right)\\
	\begin{aligned}
	\Theta_2^S(y,\beta) &= -\left(\frac{3}{8}j_0(y) + \frac{45}{56}j_2(y) + \frac{507}{448} j_4(y)\right)\\
	&+\left(\frac{1}{2}j_0(y) + \frac{5}{7}j_2(y) - \frac{81}{112} j_4(y)\right)\cos\beta\\
	&-\left(\frac{1}{8}j_0(y) - \frac{5}{56}j_2(y) + \frac{9}{448} j_4(y)\right)\cos2\beta~.
	\end{aligned}
\end{gather}

\begin{figure}
\centering
\includegraphics[keepaspectratio, scale=0.6]{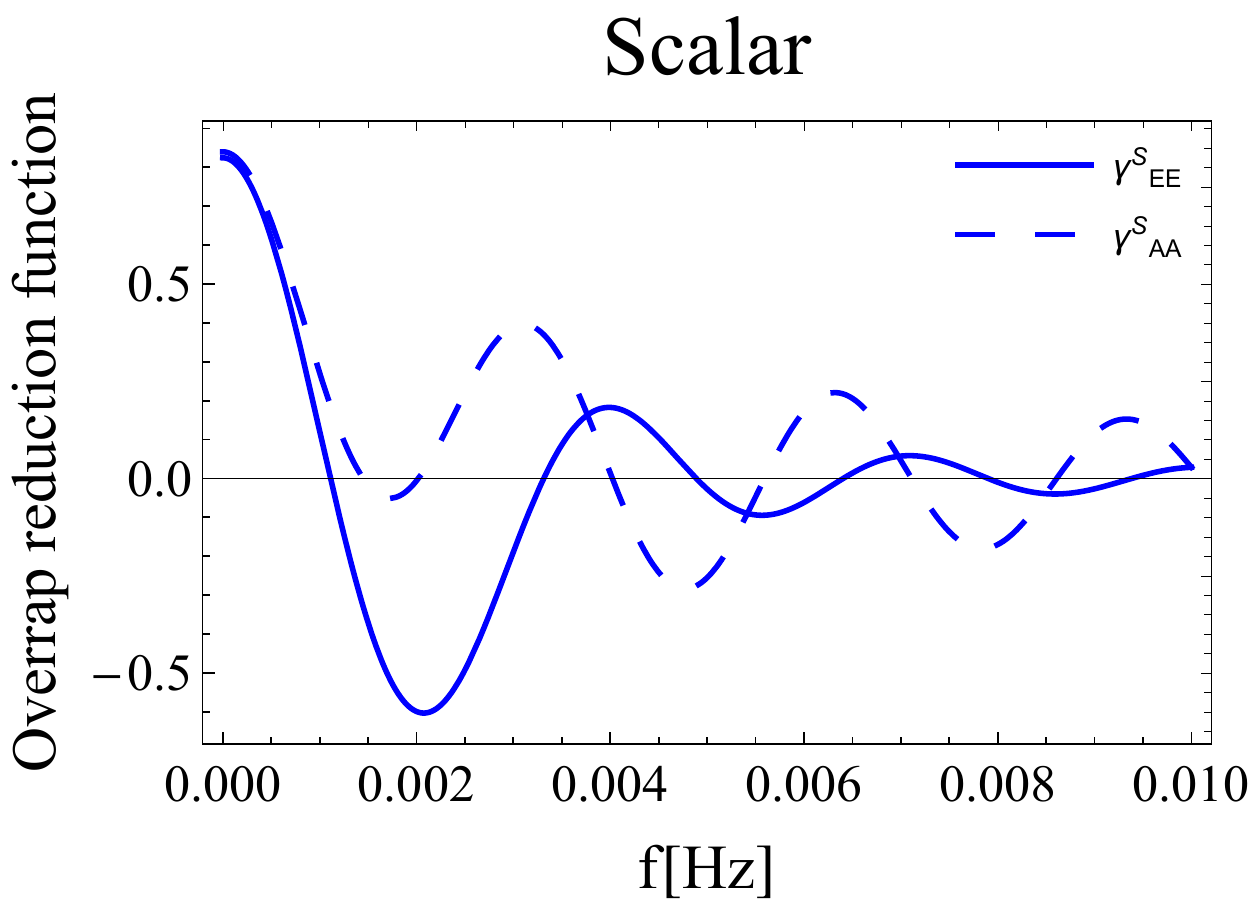}
\caption{The overlap reduction functions of the LISA-Taiji network for the scalar modes. The solid and dashed lines correspond to the $\mathrm{EE'}$ and $\mathrm{AA'}$ data combination, respectively. }
\label{fig:7}
\end{figure}

In Fig.\ref{fig:7}, we present the overlap reduction functions of the scalar modes for the $\mathrm{AA'}$ and $\mathrm{EE'}$ data pairs. Their basic profiles are qualitatively similar to $\gamma^V_{ab}(f)$  for the vector modes (see Eqs.\eqref{eq:32}-\eqref{eq:35} and the following discussion). Indeed, the function $\gamma_{ab}^S(f)$ approaches $D_{a,ij}D_{b}^{ij}/2$ at the low frequency limit $f\to 0$, and oscillates with the interval $c/d \sim 3$ mHz.

Similar to the vector modes, using Eqs.\eqref{eq:26}, \eqref{eq:27}, \eqref{eq:45}, and \eqref{eq:46}, we can evaluate the signal-to-noise ratio of the scalar modes. Its formal expression is given by
\begin{align}\label{eq:51}
\begin{aligned}
	\mathrm{SNR}_{S}^2(f_{cut}) = &\left(\frac{3 H_0^2}{10\pi^2}\right)^2 T_{obs}\\
	\times &\left[2\int_{f_{cut}}^{\infty} df \frac{\left(\Gamma^{TS}(f)\tilde{\Omega}_{GW}^S(f)\right)^2}{f^6N(f)N'(f)}\right]
\end{aligned}
\end{align}
with
\begin{align}\label{eq:52}
	\Gamma^{TS}(f) \equiv \frac{\gamma^T_{EE'}(f)\gamma^S_{AA'}(f) - \gamma^T_{AA'}(f)\gamma^S_{EE'}(f)}{\sqrt{\left(\gamma^{T}_{AA'}(f)\right)^2 + \left(\gamma^{T}_{EE'}(f)\right)^2}}~.
\end{align}
We present the effective overlap reduction function $\Gamma^{TS}(f)$ in Fig.\ref{fig:5}. In the same way as $\Gamma^{TV}(f)$, it decays rapidly in the frequency range $f \lesssim 2$ mHz, and oscillates with the frequency interval $c/2d \sim 1.5 \mathrm{mHz}$ above $f \sim 2$ mHz.

\begin{figure}[thb]
\centering
\includegraphics[keepaspectratio, scale=0.6]{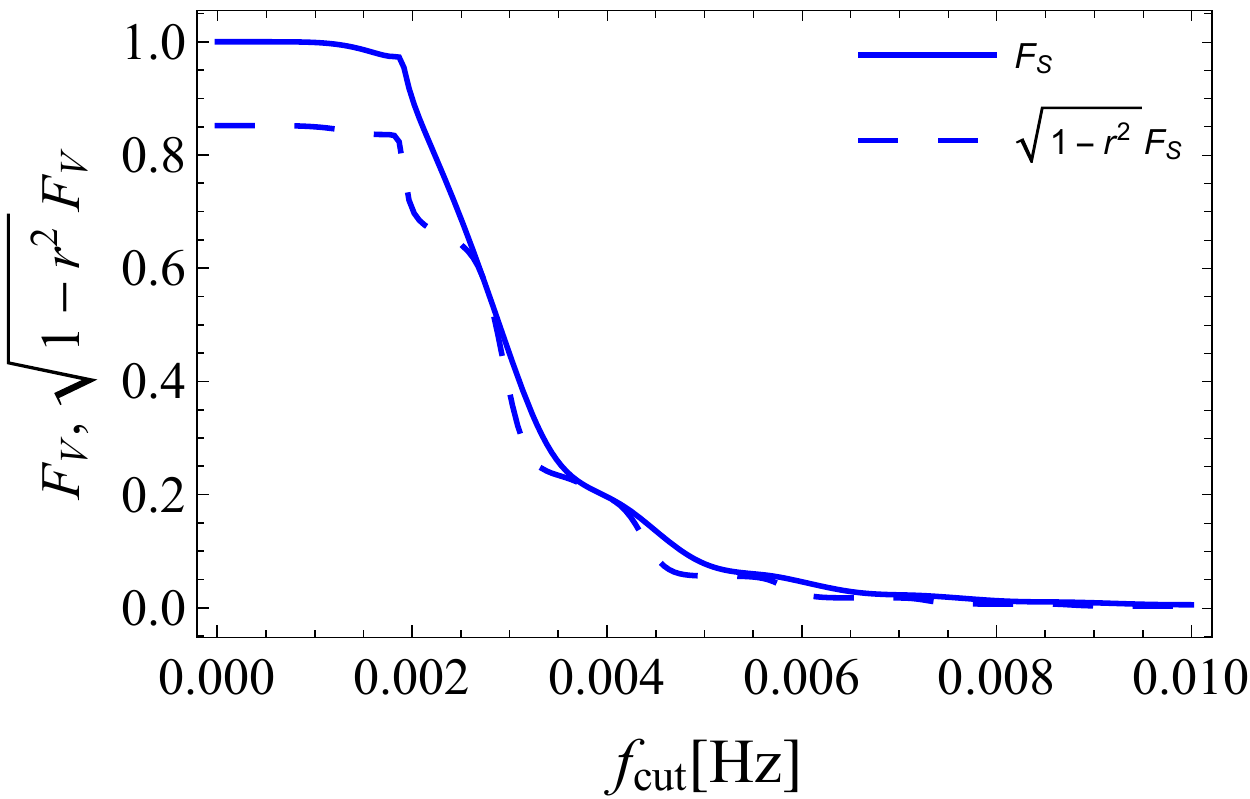}
\caption{Dependence of the statistical quantities on the frequency cutoff $f_{cut}$ for the scalar modes. The solid line shows the function $\mathcal{F}_S$ for the signal-to-noise ratio as in Eq.\eqref{eq:51}. The dashed line is for the simultaneous parameter estimation in Eq.\eqref{eq:66}.}
\label{fig:8}
\end{figure}

Now we assume the flat spectrum $\tilde{\Omega}_{GW}^S(f) = \tilde{\Omega}_{GW}^S$ for the scalar modes. Then we numerically integrate Eq.\eqref{eq:51} and obtain
\begin{align}\label{eq:53}
	\mathrm{SNR}_{S}(f_{cut}) &= 20.2\left(\frac{\tilde{\Omega}_{GW}^S}{10^{-12}}\right) \left(\frac{T_{obs}}{10 \mathrm{yr}}\right)^{1/2}\mathcal{F}_S(f_{cut})~.
\end{align}
Here the factor $\mathcal{F}_S(f_{cut})$ shows the dependence on the cut-off frequency $f_{cut}$ with the normalization
\begin{align}
	\mathcal{F}_S(0) = 1~.
\end{align}
We plot the function $\mathcal{F}_{S}(f_{cut})$ in Fig.\ref{fig:8}. Again, its overall profile is quite similar to $\mathcal{F}_{V}(f_{cut})$, presented in Fig.\ref{fig:6}. For example, the function $\mathcal{F}_{S}(f_{cut})$ depends weakly on $f_{cut}$ below 2 mHz, due to the suppression of the compiled overlap reduction function $\Gamma^{TS}(f)$ there. In addition, it has a step-like profile above 2 mHz reflecting the oscillatory feature of $\Gamma^{TV}(f)$ (but less prominent then the vector mode).

We can also estimate the error for the single fitting parameter $\tilde{\Omega}^S_{GW}$ of the flat spectrum. Similar to Eq.\eqref{eq:41},  the estimation error $\Delta\tilde{\Omega}^S_{GW}$ has a simple scaling relation \cite{Allen:1997ad, Seto:2005qy}:
\begin{align}\label{eq:55}
	\Braket{\left(\frac{\Delta \tilde{\Omega}_{GW}^S}{\tilde{\Omega}_{GW}^S}\right)^2}^{1/2} = \frac{1}{\mathrm{SNR}_S(f_{cut})}\\
	\propto \frac{1}{\mathcal{F}_S(f_{cut})}~.
\end{align}

\subsection{Simultaneous estimation of the Vector and Scalar}\label{sec:4D}

So far we have considered the vector and scalar modes separately. But, in general, the background could consist of the tensor, vector, and scalar modes at the same time. Unfortunately, with the LISA-Taiji network, we cannot further decompose the vector and scalar modes algebraically by the method described in section \ref{sec:4A} for cleaning the tensor modes. This is because the network only has two independent data pairs $\mathrm{AA'}$ and $\mathrm{EE'}$ for the parity even part, and has no freedom to isolate the three modes completely. In this section, under this restriction, we consider the parameter estimation for the two spectra $\tilde{\Omega}_{GW}^{V}(f)$ and $\tilde{\Omega}_{GW}^S(f)$ in parallel, when the background is composed by the three (T, V, and S) polarization modes. 

Our basic idea here is to use the frequency dependence of our estimator $\mu$. For the most general background, we have 
\begin{align}\label{eq:58.1}
\begin{aligned}
	C_{ab}(f) &= \frac{8\pi}{5}\left(\gamma^T_{ab}(f) S_{h}^T(f) + \gamma^V_{ab}(f) S_{h}^V(f) + \gamma^S_{ab}(f) S_{h}^S(f) \right)~.
\end{aligned}
\end{align}
Substituting Eq.\eqref{eq:58.1} to Eq.\eqref{eq:23}, we obtain
\begin{widetext}
\begin{align}
\begin{aligned}
	\braket{\mu}\bigl|_{T,V,S} &= \frac{8\pi}{5}\left[\gamma^T_{EE'}(f)\gamma^V_{AA'}(f) - \gamma^T_{AA'}(f)\gamma^V_{EE'}(f)\right]S_{h}^V(f) + \frac{8\pi}{5}\left[\gamma^T_{AA'}(f)\gamma^S_{EE'}(f) - \gamma^T_{EE'}(f)\gamma^S_{AA'}(f)\right]S_{h}^S(f)\\
	&= \frac{3 H_0^2}{10 \pi^2 f^3}\left(\left[\gamma^T_{EE'}(f)\gamma^V_{AA'}(f) - \gamma^T_{AA'}(f)\gamma^V_{EE'}(f)\right]\tilde{\Omega}_{GW}^V(f) + \left[\gamma^T_{AA'}(f)\gamma^S_{EE'}(f) - \gamma^T_{EE'}(f)\gamma^S_{AA'}(f)\right]\tilde{\Omega}_{GW}^S(f)\right)~.
\end{aligned}
\end{align}
\end{widetext}
We consider a scenario to apply the maximum likelihood analysis to our estimator $\mu$ for simultaneously fitting the two amplitudes $\tilde{\Omega}_{GW}^V$ and $\tilde{\Omega}_{GW}^S$. For simplicity, we assume that the vector and scalar modes have the flat spectra
\begin{align}
	\tilde{\Omega}_{GW}^V(f) = \tilde{\Omega}_{GW}^V~,\\
	\tilde{\Omega}_{GW}^S(f) = \tilde{\Omega}_{GW}^S~.
\end{align}
We observe that profile of the overlap reduction functions $\gamma_{AA'}^V(f), \gamma_{EE'}^V(f), \gamma_{AA'}^S(f),$ and $\gamma_{EE'}^S(f)$ induce the characteristic frequency dependence of the data combination $\mu$.

We define the error covariance matrix in the relative form:
\begin{align}
	\Sigma \equiv
	\left(
	\begin{array}{cc}
		\displaystyle \Braket{\frac{\Delta\tilde{\Omega}_{GW}^V}{\tilde{\Omega}_{GW}^V}\frac{\Delta\tilde{\Omega}_{GW}^V}{\tilde{\Omega}_{GW}^V}} & \displaystyle \Braket{\frac{\Delta\tilde{\Omega}_{GW}^V}{\tilde{\Omega}_{GW}^V}\frac{\Delta\tilde{\Omega}_{GW}^S}{\tilde{\Omega}_{GW}^S}}\\
		 & \\
		\displaystyle \Braket{\frac{\Delta\tilde{\Omega}_{GW}^S}{\tilde{\Omega}_{GW}^S}\frac{\Delta\tilde{\Omega}_{GW}^V}{\tilde{\Omega}_{GW}^V}} & \displaystyle \Braket{\frac{\Delta\tilde{\Omega}_{GW}^S}{\tilde{\Omega}_{GW}^S}\frac{\Delta\tilde{\Omega}_{GW}^S}{\tilde{\Omega}_{GW}^S}}	\end{array}
	\right)~.
\end{align}
Then, using the Fisher matrix approach \cite{Seto:2005qy}, the inverse of this matrix is given by
\begin{widetext}
\begin{align}\label{eq:61}
\begin{aligned}
	\Sigma_i^{PP'} \equiv \left(\Sigma^{-1}\right)^{PP'} = 2 T_{obs} \int_{f_{cut}}^{+\infty} df \frac{\left(\tilde{\Omega}_{GW}^P\partial_{\tilde{\Omega}_{GW}^P}\braket{\mu}\bigl|_{T,V,S}\right)\left(\tilde{\Omega}_{GW}^{P'}\partial_{\tilde{\Omega}_{GW}^{P'}}\braket{\mu}\bigl|_{T,V,S}\right)}{N(f) N'(f)}~.
\end{aligned}
\end{align}
\end{widetext}
Note that the diagonal elements are identical to $\mathrm{SNR}_V^2$ and $\mathrm{SNR}_S^2$ defined in Eqs.\eqref{eq:37} and \eqref{eq:51}
\begin{align}
	\Sigma_i^{VV} = \mathrm{SNR}_V^2~\\
	\Sigma_{i}^{SS} = \mathrm{SNR}_S^2~.
\end{align}
But the right-hand-sides of these equations do not have the original meanings of the signal-to-noise ratios as before. We keep to use these notations just for the comparison with the results for the single parameter estimations such as Eqs.\eqref{eq:39} and \eqref{eq:53}.

We define the covariance coefficient $r$ for the off-diagonal element $\Sigma_{i}^{VS}$ by
\begin{align}\label{eq:64}
	r \equiv \frac{\Sigma_{i}^{VS}}{\sqrt{\Sigma_{i}^{VV}\Sigma_{i}^{SS}}}~.
\end{align}
For the LISA-Taiji network, we can numerically evaluate the coefficient $r$ as a function of $f_{cut}$.

Now we can take the inverse of the matrix $\Sigma_i$ and obtain
\begin{widetext}
\begin{align}\label{eq:65}
		\Sigma =
	\left(
	\begin{array}{cc}
		(1-r^2)^{-1} \mathrm{SNR}_V^{-2}&	-(1-r^2)^{-1} r \ \mathrm{SNR}_V^{-1} \mathrm{SNR}_S^{-1}\\
		-(1-r^2)^{-1}r \ \mathrm{SNR}_V^{-1} \mathrm{SNR}_S^{-1} & (1-r^2)^{-1} \mathrm{SNR}_S^{-2}
	\end{array}
	\right)~.
\end{align}
\end{widetext}
Then the parameter estimation errors for the two amplitudes (P = V and S) are given by
\begin{align}\label{eq:66}
	\Braket{\left(\frac{\Delta\tilde{\Omega}_{GW}^P}{\tilde{\Omega}_{GW}^P}\right)^2}^{1/2} &= \frac{1}{\sqrt{1- r^2}}\frac{1}{\mathrm{SNR}_{P}}\\
	&\propto \frac{1}{\sqrt{1-r(f_{cut})^2}}\frac{1}{\mathcal{F}_P(f_{cut})}~.
\end{align}
We should compare Eq.\eqref{eq:66} directly with Eqs.\eqref{eq:41} and \eqref{eq:55} for the single parameter estimation. In this expression, the factor $(1-r^2)^{-1/2} (>1)$ presents the increment of the errors associated with noise covariance of the two parameter fitting, compared with the single parameter estimation. In addition, as shown in Eq.\eqref{eq:65}, the covariance coefficient of the error is given by $-r$.

In Figs.\ref{fig:5} and \ref{fig:8}, we present the products $\sqrt{1-r^2}\mathcal{F}_{P}$ $(P = V,$ and $S)$ as  functions of the low frequency cut-off $f_{cut}$. The statistical loss $\sqrt{1-r^2}$ is $\sim 0.2$ for $f_{cut} \lesssim 2$ mHz. Also,  at some frequencies, we have $\sqrt{1-r^2}\mathcal{F}_P = \mathcal{F}_P$, corresponding to $r=0$. This is due to  the oscillations of the overlap reduction functions. In general, we have $r\sim 1$ when the effective dynamic range of the frequency integral decreases. Using Figs.\ref{fig:5} and \ref{fig:8}, together with Eqs.\eqref{eq:39} and \eqref{eq:53}, we can evaluate the actual expectation  values for the parameter estimation errors in our flat spectral model.

\section{Other network geometries}\label{sec:4.5}

So far, we have examined the fixed network geometry characterized by the orbital phase difference $\Delta \theta=40^\circ$ (equivalently the opening angle $\beta=34.46^\circ$), as shown in Fig.1.  But, the orbital designs of LISA and Taiji have not been finalized yet. It would be thus beneficial to discuss the prospects for other potential configurations.

In this section, we first examine the networks with various phase angles $\Delta \theta$, still keeping the geometrical symmetry characterized by the virtual contact sphere (Sec. V.A).  Then, in Sec. V.B, we consider general network geometry without the virtual contact sphere. We 
clarify the conditions with which  we cannot algebraically decompose the tensor, vector, and scalar modes.

\subsection{Orbital Phase Difference}\label{sec:4.5A}

We now examine how the network sensitivities ${\rm SNR}_V$ and ${\rm SNR}_S$ depend on the orbital phase difference $\Delta \theta$.  Note that, the geometrical symmetry of the network still prohibits the algebraic decomposition of the vector and scalar modes. For simplicity, we fix the lower cut-off frequency at $f_{cut}=2$mHz. In the top panel of Fig.9, we present our numerical results. 
Around $\Delta \theta=40^\circ$, the function ${\rm SNR}_S$ is close the globally maximum value, but  ${\rm SNR}_V$ is $\sim30\%$ smaller than   the peak value around $\Delta\theta\sim 28^\circ$.    
At $\Delta\theta=0$, the overlap reduction functions of the three polarization modes are totally degenerated with $\gamma_{ab}^T=\gamma_{ab}^V=\gamma_{ab}^S$,  and  we lost sensitivities  to the vector and scalar modes (namely ${\rm SNR}_V={\rm SNR}_S=0$),  after subtracting the tensor modes. 

In the bottom panel of Fig.9, we show the covariance coefficient $r$ in the form $\sqrt{1-r^2}$. Because of the sharp frequency cut-off at $f_{cut}=2$mHz and the wavy profiles of the overlap reduction functions, the curve shows a complicated shape.

\begin{figure}
\centering
\includegraphics[keepaspectratio, scale=0.6]{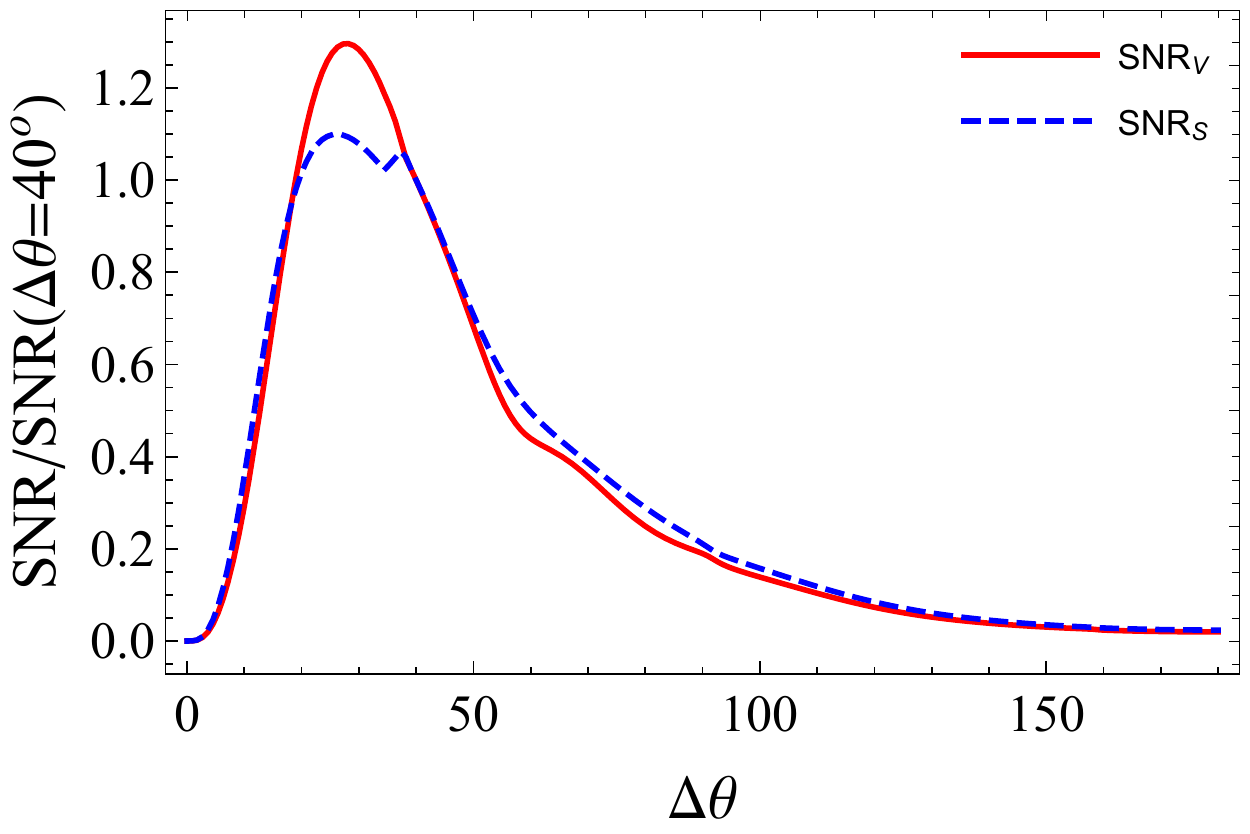}
\includegraphics[keepaspectratio, scale=0.6]{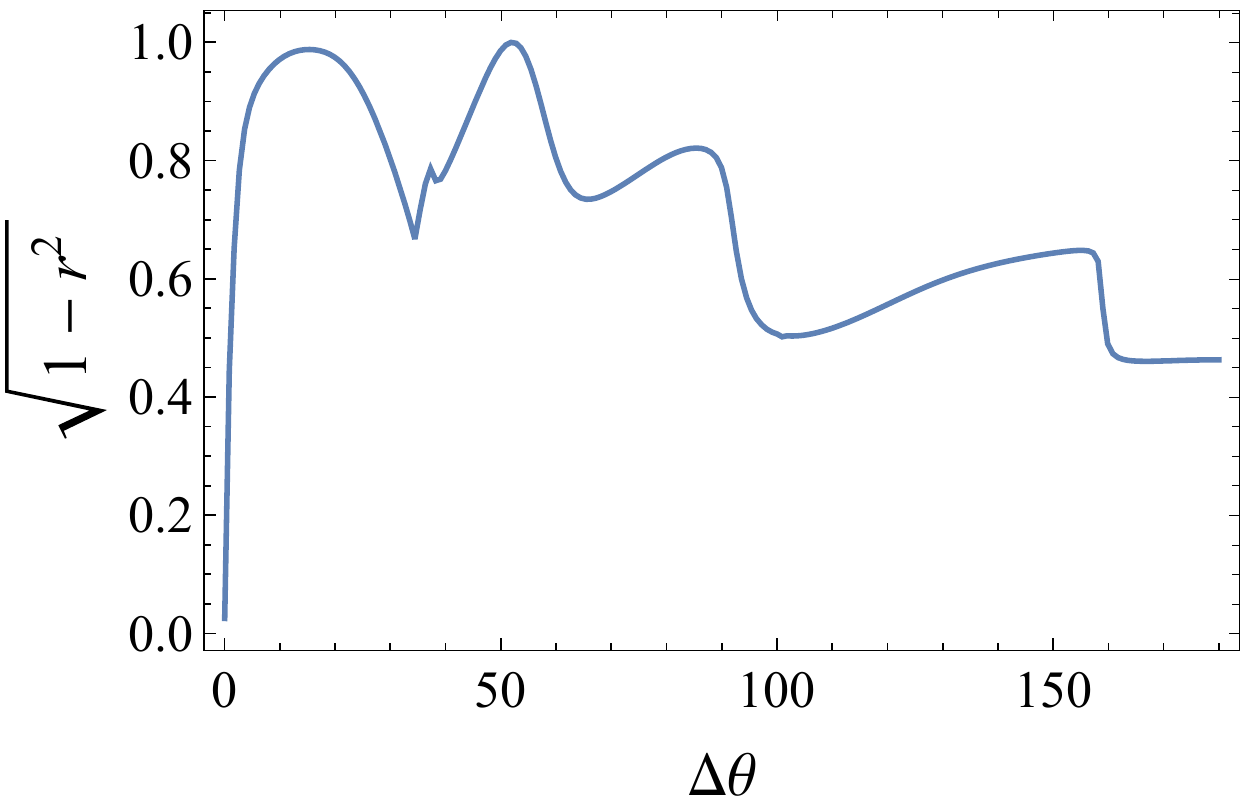}
\caption{Dependence of various statistical quantities on the orbital phase difference $\Delta \theta$. (TOP) The red line and blue dashed line show the signal-to-noise ratios of the vector and scalar modes after removing the tensor modes (see Eqs.(\ref{eq:37}) and (\ref{eq:51})).  We normalized the signal-to-noise ratios by the results at $\Delta \theta = 40^\circ$. (Bottom) The magnitude of the covariance coefficient $r$ in the form $\sqrt{1-r^2}$.}
\label{fig:9}
\end{figure}

\subsection{General Configuration}\label{sec:4.5B}

Next we consider  a general network geometry for two triangular detectors. We can formally write down the equation for the three spectra as
\begin{align}\label{eq:70}
	\left(
	\begin{array}{c}
		C_{{\rm AA'}}\\
		C_{{\rm EE'}}\\
		C_{{\rm AE'}}\\
		C_{{\rm EA'}}
	\end{array}
	\right) = 
		\frac{8\pi}{5} \mathcal{M}
		\left(
	\begin{array}{c}
		\displaystyle S_{h}^T\\
		\displaystyle S_{h}^V\\
		\displaystyle S_{h}^S
	\end{array}
	\right)~
\end{align}
with the following matrix determined by the overlap reduction functions
\begin{align}
	\mathcal{M} \equiv \left(
	\begin{array}{ccc}
		\displaystyle \gamma^T_{{\rm AA'}} & \gamma^V_{{\rm AA'}} & \gamma^S_{{\rm AA'}}\\
		\displaystyle \gamma^T_{{\rm EE'}} & \gamma^V_{{\rm EE'}} & \gamma^S_{{\rm EE'}}\\
		\displaystyle \gamma^T_{{\rm AE'}} & \gamma^V_{{\rm AE'}} & \gamma^S_{{\rm AE'}}\\
		\displaystyle \gamma^T_{{\rm EA'}} & \gamma^V_{{\rm EA'}} & \gamma^S_{{\rm EA'}}
	\end{array}
	\right)
\end{align}
(see Eq.\eqref{eq:58.1}). Under the presence of the virtual contact sphere, using a mirror symmetry, we can take $\gamma_{AE\rq{}}^P=\gamma_{EA\rq{}}^P=0$ (for $P=T,V$ and $S$), and we cannot separately solve the three spectra.  This can be attributed  to the insufficient rank of the matrix $\mathcal M$. We should notice that the rank of the matrix $\mathcal M$ is not affected by the detuning of the alignment angle $\phi$ in Fig.2, since the resultant overlap reduction functions are given by  simple linear combinations of the original aligned ones.

In any case, the three spectra can be fully separated, if the rank  of the matrix  $\mathcal M$ is three.  Using the basic tensorial expressions (see Eq.(10) of \cite{Nishizawa:2009jh}) for the overlap reduction functions, we found that the matrix  $\mathcal M$ is factorized into two matrices as  $\mathcal M=M_1\cdot M_2$. Here $M_1$ is a $3\times3$ matrix whose  components are given by linear combinations of the three Bessel functions $j_i(y)$ ($i=0,2$ and 4)  with $y=2\pi f d/c$.  We also have $\det [M_1]\propto j_0(y)j_2(y)j_4(y)$.

The second matrix $M_2$ is a $3\times 4$ matrix and independent of the parameter $y$. Its elements are given by   the angular parameters of the network formed by triangular detectors $a$ and $b$. Except for the discrete frequencies at the zero points of the product $j_0(y)j_2(y)j_4(y)$, the rank of the matrix $\mathcal M$ is determined by that of $M_2$.    To be concrete, we introduce the three unit vectors $\bm{n}_a$, $\bm{n}_b$ and $\bm{m}$. Here $\bm{n}_a$ and $\bm{n}_b$ are normal to the two detector planes,  and $\bm m$ is the unit  directional vector connecting two detectors. After some algebra, we found that the rank of $M_2$ is less than three, when one of the following two conditions is satisfied;\\
(i) The normal vectors $\bm{n}_a$ and  $\bm{n}_b$ are both orthogonal to $\bm{m}$.\\
(ii) The three vectors, $\bm{m},\bm{n}_a$ and $\bm{n}_b$ are on the same plane.\\
Below, using these simple criteria, we qualitatively discuss the possibility of the algebraic decomposition for various potential networks.

The network geometry in Fig.1 (and its variations in the previous subsection) meets the condition (ii) and we cannot make the full decomposition, as already discussed. 

Actually, in Fig.1, we can take the mirror image of each triangle with respect to the ecliptic plane. The resultant triangle can be still composed by three solutions of heliocentric orbits. Here we consider a network formed by the mirrored Taiji and the unchanged LISA. This twisted network does not satisfy the two conditions, and we can make the algebraic separation of the three spectra.   

Next, if the semi-major axises of LISA and  Taiji are different,  the two conditions are not generally satisfied.  Moreover, in this case, the matrix $M_1$ changes with time, due to the drift of the mutual distance $d$. Then the singular  frequencies corresponding to $j_0(y)j_2(y)j_4(y)=0$ also change with time.  As a result, in contrast  to a network with a fixed distance $d$, we can also dissolve the singular frequencies.

We have focused our attention to networks  formed by heliocentric detectors such as the LISA-Taiji pair and its variations. 
 We should notice that TianQin will have a geocentric orbit and its detector plane will change with time, relative to LISA.
Therefore, in  most of their operation time, the LISA-TianQin network does not satisfy the two conditions and allows us to make the algebraic decomposition.

\section{Summary and Discussion}\label{sec:5}
In this paper, we discussed a search for the vector and scalar polarization modes of isotropic stochastic gravitational wave background around 1-10 mHz with the LISA-Taiji detector network. These modes do not appear in GR, and their measurement allows us to observationally study theories of gravity. 

Because of the underlying symmetries of the network, for the even parity components, we can use two independent correlation products from the pairs AA$'$ and EE$'$. By taking their appropriate combination $\mu$, defined in Eq.\eqref{eq:22}, we can algebraically cancel the contribution of the tensor modes and examine the existence of the vector and scalar modes in a model independent way.

To clarify our basic idea, we assumed that the vector and scalar modes have flat spectra in terms of the effective energy densities $\tilde{\Omega}_{GW}^V$ and $\tilde{\Omega}_{GW}^S$ defined in Eqs.\eqref{eq:30} and \eqref{eq:45}. We first studied the case when we only have the vector modes (Sec.\ref{sec:4B}) or the scalar modes (Sec.\ref{sec:4C}), other than the tensor modes. We found that after ten years observation, the detection limit could reach $\tilde{\Omega}_{GW}^V \sim 10^{-12}$ and $\tilde{\Omega}_{GW}^S \sim 10^{-12}$. These limits are much smaller than the current upper bound $\tilde{\Omega}_{GW}^V \lesssim 1.2\times10^{-7}$ and $\tilde{\Omega}_{GW}^S \lesssim 4.2\times10^{-7}$ around 10 - 100 Hz with the ground based detectors \cite{LIGOScientific:2019vic}.

Similarly  to \cite{seto:xxxx}, we have  paid special attention to the impact of the low frequency cut off $f_{cut}$ on the accumulation of the signal-to-noise ratios. The actual value of $f_{cut}$ would be determined by the subtraction of the Galactic binary foreground and would be closely related to the operation periods of the detectors. As shown in Figs.\ref{fig:5} and \ref{fig:7}, we found that the signal-to-noise ratios depend strongly on $f_{cut} \gtrsim 2$ mHz, but weakly on $f_{cut} \lesssim 2$ mHz due to the degeneracy of the overlap reduction functions $\gamma^T_{ab} \sim \gamma^V_{ab} \sim \gamma^S_{ab}$ there. These results might be interesting when planning possible collaboration between LISA and Taiji.

Then, we considered the general case in which a background is composed of the tensor, vector, and scalar modes all together. An algebraic decomposition of all the three modes is not possible, because we need at least three correlation outputs. But, using the frequency dependence of the overlap reduction functions, we can simultaneously fit the parameters of both the vector and scalar spectra from our estimator $\mu$.  As a demonstration, we considered a situation to make the standard maximum likelihood analysis to our estimator $\mu$.  Applying the Fisher matrix formalism to the amplitudes $\tilde{\Omega}_{GW}^V$ and $\tilde{\Omega}_{GW}^S$ of our flat spectra, we evaluated their estimation errors. In this case, the covariance coefficient $r$ is the key quantity. For $f_{cut} \lesssim 2$ mHz, the estimation errors are $\sim 20 \%$ larger than the simplified cases without the blending of the vector and scalar modes.

Given the current design of the LISA-Taiji network, we have focused our attention on the specific network geometry with the orbital phase difference $\Delta \theta=40^\circ$.  But, 
in Sec.V, we discussed the prospects for other network configurations.  In Sec.\ref{sec:4.5A}, we changed the orbital angle $\Delta \theta$, keeping the virtual contact sphere.  We found that  the current design  $\Delta\theta=40^\circ$ is within $15^\circ$ of the optimal choices for ${\rm SNR}_V$ and ${\rm SNR}_S$,  as shown in Fig.9.

Because of the mirror symmetry, the contact sphere allows us to decompose the odd and even parity components of an isotropic gravitational wave background clearly \cite{seto:xxxx}.  But, at the same time, the symmetry prohibits us from algebraically decomposing the tensor, vector and scalar modes of even parity. In Sec.\ref{sec:4.5B}, we clarify the geometric conditions (i) and (ii) for the  impossibility of the full mode decomposition.  They would be useful for designing network geometry from the viewpoints of the anomalous polarization search. 

\begin{acknowledgments}
 This work is supported by JSPS Kakenhi Grant-in-Aid for Scientific Research
 (Nos. 17H06358 and 19K03870).
\end{acknowledgments}

\appendix*
\section{effective energy densities and strain fluctuations}\label{App:A}

In Eqs.\eqref{eq:3}, \eqref{eq:28}, and \eqref{eq:43}, we set the normalization of the polarization tensors to have
\begin{align}
	\gamma_{aa}^T = \gamma_{aa}^V  =  \gamma_{aa}^S=1
\end{align}
for the self-correlation of a single L-shaped interferometer. Then, together with our definitions of $S_{h}^P$ and $\tilde{\Omega}_{GW}^P \ (P = T,V,S)$, we obtain
\begin{align}\label{eq:A2}
	\braket{h_a(f)h^*_{a}(f)} &= \frac{8 \pi}{5}\left(S_h^T(f) + S_h^V(f) + S_h^S(f)\right)\\
	\label{eq:A3}
	&= \frac{3 H_0^2}{10 \pi^2 f^3}\left(\Omega_{GW}^T(f) + \tilde{\Omega}_{GW}^V(f) + \tilde{\Omega}_{GW}^S(f)\right).
\end{align}
In fact, for the vector and scalar modes, we fix their polarization tensors, power spectra and effective energy densities, to realize the organized forms \eqref{eq:A2} and \eqref{eq:A3} for the strain fluctuations induced by the three polarization modes. In this paper we do not deal with the actual energy densities that depend on the details of the gravity theories \cite{Isi:2018miq}.

\bibliography{ref}

\providecommand{\noopsort}[1]{}\providecommand{\singleletter}[1]{#1}%
\begin{thebibliography}{31}
\expandafter\ifx\csname natexlab\endcsname\relax\def\natexlab#1{#1}\fi
\expandafter\ifx\csname bibnamefont\endcsname\relax
  \def\bibnamefont#1{#1}\fi
\expandafter\ifx\csname bibfnamefont\endcsname\relax
  \def\bibfnamefont#1{#1}\fi
\expandafter\ifx\csname citenamefont\endcsname\relax
  \def\citenamefont#1{#1}\fi
\expandafter\ifx\csname url\endcsname\relax
  \def\url#1{\texttt{#1}}\fi
\expandafter\ifx\csname urlprefix\endcsname\relax\def\urlprefix{URL }\fi
\providecommand{\bibinfo}[2]{#2}
\providecommand{\eprint}[2][]{\url{#2}}

\bibitem[{\citenamefont{Starobinsky}(1979)}]{Starobinsky:1979ty}
\bibinfo{author}{\bibfnamefont{A.~A.} \bibnamefont{Starobinsky}},
  \bibinfo{journal}{JETP Lett.} \textbf{\bibinfo{volume}{30}},
  \bibinfo{pages}{682} (\bibinfo{year}{1979}).

\bibitem[{\citenamefont{Easther et~al.}(2007)\citenamefont{Easther, Giblin, and
  Lim}}]{PhysRevLett.99.221301}
\bibinfo{author}{\bibfnamefont{R.}~\bibnamefont{Easther}},
  \bibinfo{author}{\bibfnamefont{J.~T.} \bibnamefont{Giblin}},
  \bibnamefont{and} \bibinfo{author}{\bibfnamefont{E.~A.} \bibnamefont{Lim}},
  \bibinfo{journal}{Phys. Rev. Lett.} \textbf{\bibinfo{volume}{99}},
  \bibinfo{pages}{221301} (\bibinfo{year}{2007}).

\bibitem[{\citenamefont{Cook and Sorbo}(2012)}]{Cook:2011hg}
\bibinfo{author}{\bibfnamefont{J.~L.} \bibnamefont{Cook}} \bibnamefont{and}
  \bibinfo{author}{\bibfnamefont{L.}~\bibnamefont{Sorbo}},
  \bibinfo{journal}{Phys. Rev. D} \textbf{\bibinfo{volume}{85}},
  \bibinfo{pages}{023534} (\bibinfo{year}{2012}), \bibinfo{note}{[Erratum:
  Phys.Rev.D 86, 069901 (2012)]}, \eprint{1109.0022}.

\bibitem[{\citenamefont{Kamionkowski et~al.}(1994)\citenamefont{Kamionkowski,
  Kosowsky, and Turner}}]{Kamionkowski:1993fg}
\bibinfo{author}{\bibfnamefont{M.}~\bibnamefont{Kamionkowski}},
  \bibinfo{author}{\bibfnamefont{A.}~\bibnamefont{Kosowsky}}, \bibnamefont{and}
  \bibinfo{author}{\bibfnamefont{M.~S.} \bibnamefont{Turner}},
  \bibinfo{journal}{Phys. Rev. D} \textbf{\bibinfo{volume}{49}},
  \bibinfo{pages}{2837} (\bibinfo{year}{1994}), \eprint{astro-ph/9310044}.

\bibitem[{\citenamefont{Caprini et~al.}(2008)\citenamefont{Caprini, Durrer, and
  Servant}}]{Caprini:2007xq}
\bibinfo{author}{\bibfnamefont{C.}~\bibnamefont{Caprini}},
  \bibinfo{author}{\bibfnamefont{R.}~\bibnamefont{Durrer}}, \bibnamefont{and}
  \bibinfo{author}{\bibfnamefont{G.}~\bibnamefont{Servant}},
  \bibinfo{journal}{Phys. Rev. D} \textbf{\bibinfo{volume}{77}},
  \bibinfo{pages}{124015} (\bibinfo{year}{2008}), \eprint{0711.2593}.

\bibitem[{\citenamefont{Maggiore}(2000)}]{Maggiore:1999vm}
\bibinfo{author}{\bibfnamefont{M.}~\bibnamefont{Maggiore}},
  \bibinfo{journal}{Phys. Rept.} \textbf{\bibinfo{volume}{331}},
  \bibinfo{pages}{283} (\bibinfo{year}{2000}), \eprint{gr-qc/9909001}.

\bibitem[{\citenamefont{Damour and Vilenkin}(2005)}]{Damour:2004kw}
\bibinfo{author}{\bibfnamefont{T.}~\bibnamefont{Damour}} \bibnamefont{and}
  \bibinfo{author}{\bibfnamefont{A.}~\bibnamefont{Vilenkin}},
  \bibinfo{journal}{Phys. Rev. D} \textbf{\bibinfo{volume}{71}},
  \bibinfo{pages}{063510} (\bibinfo{year}{2005}), \eprint{hep-th/0410222}.

\bibitem[{\citenamefont{Olmez et~al.}(2010)\citenamefont{Olmez, Mandic, and
  Siemens}}]{Olmez:2010bi}
\bibinfo{author}{\bibfnamefont{S.}~\bibnamefont{Olmez}},
  \bibinfo{author}{\bibfnamefont{V.}~\bibnamefont{Mandic}}, \bibnamefont{and}
  \bibinfo{author}{\bibfnamefont{X.}~\bibnamefont{Siemens}},
  \bibinfo{journal}{Phys. Rev. D} \textbf{\bibinfo{volume}{81}},
  \bibinfo{pages}{104028} (\bibinfo{year}{2010}), \eprint{1004.0890}.

\bibitem[{\citenamefont{Farmer and Phinney}(2003)}]{Farmer:2003pa}
\bibinfo{author}{\bibfnamefont{A.~J.} \bibnamefont{Farmer}} \bibnamefont{and}
  \bibinfo{author}{\bibfnamefont{E.}~\bibnamefont{Phinney}},
  \bibinfo{journal}{Mon. Not. Roy. Astron. Soc.}
  \textbf{\bibinfo{volume}{346}}, \bibinfo{pages}{1197} (\bibinfo{year}{2003}),
  \eprint{astro-ph/0304393}.

\bibitem[{\citenamefont{Marassi et~al.}(2011)\citenamefont{Marassi, Schneider,
  Corvino, Ferrari, and Zwart}}]{PhysRevD.84.124037}
\bibinfo{author}{\bibfnamefont{S.}~\bibnamefont{Marassi}},
  \bibinfo{author}{\bibfnamefont{R.}~\bibnamefont{Schneider}},
  \bibinfo{author}{\bibfnamefont{G.}~\bibnamefont{Corvino}},
  \bibinfo{author}{\bibfnamefont{V.}~\bibnamefont{Ferrari}}, \bibnamefont{and}
  \bibinfo{author}{\bibfnamefont{S.~P.} \bibnamefont{Zwart}},
  \bibinfo{journal}{Phys. Rev. D} \textbf{\bibinfo{volume}{84}},
  \bibinfo{pages}{124037} (\bibinfo{year}{2011}).

\bibitem[{\citenamefont{Zhu et~al.}(2013)\citenamefont{Zhu, Howell, Blair, and
  Zhu}}]{Zhu:2012xw}
\bibinfo{author}{\bibfnamefont{X.-J.} \bibnamefont{Zhu}},
  \bibinfo{author}{\bibfnamefont{E.~J.} \bibnamefont{Howell}},
  \bibinfo{author}{\bibfnamefont{D.~G.} \bibnamefont{Blair}}, \bibnamefont{and}
  \bibinfo{author}{\bibfnamefont{Z.-H.} \bibnamefont{Zhu}},
  \bibinfo{journal}{Mon. Not. Roy. Astron. Soc.}
  \textbf{\bibinfo{volume}{431}}, \bibinfo{pages}{882} (\bibinfo{year}{2013}),
  \eprint{1209.0595}.

\bibitem[{\citenamefont{Christensen}(2019)}]{Christensen:2018iqi}
\bibinfo{author}{\bibfnamefont{N.}~\bibnamefont{Christensen}},
  \bibinfo{journal}{Rept. Prog. Phys.} \textbf{\bibinfo{volume}{82}},
  \bibinfo{pages}{016903} (\bibinfo{year}{2019}), \eprint{1811.08797}.

\bibitem[{\citenamefont{Will}(1993)}]{Will:1993ns}
\bibinfo{author}{\bibfnamefont{C.}~\bibnamefont{Will}},
  \emph{\bibinfo{title}{{Theory and experiment in gravitational physics}}}
  (\bibinfo{year}{1993}), ISBN \bibinfo{isbn}{978-0-521-43973-2}.

\bibitem[{\citenamefont{Flanagan}(1993)}]{Flanagan:1993ix}
\bibinfo{author}{\bibfnamefont{E.~E.} \bibnamefont{Flanagan}},
  \bibinfo{journal}{Phys. Rev. D} \textbf{\bibinfo{volume}{48}},
  \bibinfo{pages}{2389} (\bibinfo{year}{1993}), \eprint{astro-ph/9305029}.

\bibitem[{\citenamefont{Allen and Romano}(1999)}]{Allen:1997ad}
\bibinfo{author}{\bibfnamefont{B.}~\bibnamefont{Allen}} \bibnamefont{and}
  \bibinfo{author}{\bibfnamefont{J.~D.} \bibnamefont{Romano}},
  \bibinfo{journal}{Phys. Rev. D} \textbf{\bibinfo{volume}{59}},
  \bibinfo{pages}{102001} (\bibinfo{year}{1999}), \eprint{gr-qc/9710117}.

\bibitem[{\citenamefont{Nishizawa et~al.}(2009)\citenamefont{Nishizawa, Taruya,
  Hayama, Kawamura, and Sakagami}}]{Nishizawa:2009bf}
\bibinfo{author}{\bibfnamefont{A.}~\bibnamefont{Nishizawa}},
  \bibinfo{author}{\bibfnamefont{A.}~\bibnamefont{Taruya}},
  \bibinfo{author}{\bibfnamefont{K.}~\bibnamefont{Hayama}},
  \bibinfo{author}{\bibfnamefont{S.}~\bibnamefont{Kawamura}}, \bibnamefont{and}
  \bibinfo{author}{\bibfnamefont{M.-a.} \bibnamefont{Sakagami}},
  \bibinfo{journal}{Phys. Rev. D} \textbf{\bibinfo{volume}{79}},
  \bibinfo{pages}{082002} (\bibinfo{year}{2009}), \eprint{0903.0528}.

\bibitem[{\citenamefont{Nishizawa et~al.}(2010)\citenamefont{Nishizawa, Taruya,
  and Kawamura}}]{Nishizawa:2009jh}
\bibinfo{author}{\bibfnamefont{A.}~\bibnamefont{Nishizawa}},
  \bibinfo{author}{\bibfnamefont{A.}~\bibnamefont{Taruya}}, \bibnamefont{and}
  \bibinfo{author}{\bibfnamefont{S.}~\bibnamefont{Kawamura}},
  \bibinfo{journal}{Phys. Rev. D} \textbf{\bibinfo{volume}{81}},
  \bibinfo{pages}{104043} (\bibinfo{year}{2010}), \eprint{0911.0525}.

\bibitem[{\citenamefont{Abbott et~al.}(2019)}]{LIGOScientific:2019vic}
\bibinfo{author}{\bibfnamefont{B.}~\bibnamefont{Abbott}} \bibnamefont{et~al.}
  (\bibinfo{collaboration}{LIGO Scientific, Virgo}), \bibinfo{journal}{Phys.
  Rev. D} \textbf{\bibinfo{volume}{100}}, \bibinfo{pages}{061101}
  (\bibinfo{year}{2019}), \eprint{1903.02886}.

\bibitem[{\citenamefont{Cornish et~al.}(2018)\citenamefont{Cornish, O'Beirne,
  Taylor, and Yunes}}]{Cornish:2017oic}
\bibinfo{author}{\bibfnamefont{N.~J.} \bibnamefont{Cornish}},
  \bibinfo{author}{\bibfnamefont{L.}~\bibnamefont{O'Beirne}},
  \bibinfo{author}{\bibfnamefont{S.~R.} \bibnamefont{Taylor}},
  \bibnamefont{and} \bibinfo{author}{\bibfnamefont{N.}~\bibnamefont{Yunes}},
  \bibinfo{journal}{Phys. Rev. Lett.} \textbf{\bibinfo{volume}{120}},
  \bibinfo{pages}{181101} (\bibinfo{year}{2018}), \eprint{1712.07132}.

\bibitem[{\citenamefont{Amaro-Seoane et~al.}(2017)}]{Audley:2017drz}
\bibinfo{author}{\bibfnamefont{P.}~\bibnamefont{Amaro-Seoane}}
  \bibnamefont{et~al.} (\bibinfo{collaboration}{LISA}) (\bibinfo{year}{2017}),
  \eprint{1702.00786}.

\bibitem[{\citenamefont{Hu and Wu}(2017)}]{Hu:2017mde}
\bibinfo{author}{\bibfnamefont{W.-R.} \bibnamefont{Hu}} \bibnamefont{and}
  \bibinfo{author}{\bibfnamefont{Y.-L.} \bibnamefont{Wu}},
  \bibinfo{journal}{Natl. Sci. Rev.} \textbf{\bibinfo{volume}{4}},
  \bibinfo{pages}{685} (\bibinfo{year}{2017}).

\bibitem[{\citenamefont{Luo et~al.}(2016)}]{Luo:2015ght}
\bibinfo{author}{\bibfnamefont{J.}~\bibnamefont{Luo}} \bibnamefont{et~al.}
  (\bibinfo{collaboration}{TianQin}), \bibinfo{journal}{Class. Quant. Grav.}
  \textbf{\bibinfo{volume}{33}}, \bibinfo{pages}{035010}
  (\bibinfo{year}{2016}), \eprint{1512.02076}.

\bibitem[{\citenamefont{Seto}(2004)}]{Seto:2004ji}
\bibinfo{author}{\bibfnamefont{N.}~\bibnamefont{Seto}}, \bibinfo{journal}{Phys.
  Rev. D} \textbf{\bibinfo{volume}{69}}, \bibinfo{pages}{123005}
  (\bibinfo{year}{2004}), \eprint{gr-qc/0403014}.

\bibitem[{\citenamefont{Seto}(2020)}]{seto:xxxx}
\bibinfo{author}{\bibfnamefont{N.}~\bibnamefont{Seto}} (\bibinfo{year}{2020}),
  \eprint{2009.02928}.

\bibitem[{\citenamefont{Seto and Taruya}(2008)}]{Seto:2008sr}
\bibinfo{author}{\bibfnamefont{N.}~\bibnamefont{Seto}} \bibnamefont{and}
  \bibinfo{author}{\bibfnamefont{A.}~\bibnamefont{Taruya}},
  \bibinfo{journal}{Phys. Rev. D} \textbf{\bibinfo{volume}{77}},
  \bibinfo{pages}{103001} (\bibinfo{year}{2008}), \eprint{0801.4185}.

\bibitem[{\citenamefont{Prince et~al.}(2002)\citenamefont{Prince, Tinto,
  Larson, and Armstrong}}]{PhysRevD.66.122002}
\bibinfo{author}{\bibfnamefont{T.~A.} \bibnamefont{Prince}},
  \bibinfo{author}{\bibfnamefont{M.}~\bibnamefont{Tinto}},
  \bibinfo{author}{\bibfnamefont{S.~L.} \bibnamefont{Larson}},
  \bibnamefont{and} \bibinfo{author}{\bibfnamefont{J.~W.}
  \bibnamefont{Armstrong}}, \bibinfo{journal}{Phys. Rev. D}
  \textbf{\bibinfo{volume}{66}}, \bibinfo{pages}{122002}
  (\bibinfo{year}{2002}).

\bibitem[{\citenamefont{Isi and Stein}(2018)}]{Isi:2018miq}
\bibinfo{author}{\bibfnamefont{M.}~\bibnamefont{Isi}} \bibnamefont{and}
  \bibinfo{author}{\bibfnamefont{L.~C.} \bibnamefont{Stein}},
  \bibinfo{journal}{Phys. Rev. D} \textbf{\bibinfo{volume}{98}},
  \bibinfo{pages}{104025} (\bibinfo{year}{2018}), \eprint{1807.02123}.

\bibitem[{\citenamefont{Robson et~al.}(2019)\citenamefont{Robson, Cornish, and
  Liu}}]{Cornish:2018dyw}
\bibinfo{author}{\bibfnamefont{T.}~\bibnamefont{Robson}},
  \bibinfo{author}{\bibfnamefont{N.~J.} \bibnamefont{Cornish}},
  \bibnamefont{and} \bibinfo{author}{\bibfnamefont{C.}~\bibnamefont{Liu}},
  \bibinfo{journal}{Class. Quant. Grav.} \textbf{\bibinfo{volume}{36}},
  \bibinfo{pages}{105011} (\bibinfo{year}{2019}), \eprint{1803.01944}.

\bibitem[{\citenamefont{Wang et~al.}(2020)\citenamefont{Wang, Ni, Han, Yang,
  and Zhong}}]{Wang:2020vkg}
\bibinfo{author}{\bibfnamefont{G.}~\bibnamefont{Wang}},
  \bibinfo{author}{\bibfnamefont{W.-T.} \bibnamefont{Ni}},
  \bibinfo{author}{\bibfnamefont{W.-B.} \bibnamefont{Han}},
  \bibinfo{author}{\bibfnamefont{S.-C.} \bibnamefont{Yang}}, \bibnamefont{and}
  \bibinfo{author}{\bibfnamefont{X.-Y.} \bibnamefont{Zhong}}
  (\bibinfo{year}{2020}), \eprint{2002.12628}.

\bibitem[{\citenamefont{Seto}(2006)}]{Seto:2005qy}
\bibinfo{author}{\bibfnamefont{N.}~\bibnamefont{Seto}}, \bibinfo{journal}{Phys.
  Rev. D} \textbf{\bibinfo{volume}{73}}, \bibinfo{pages}{063001}
  (\bibinfo{year}{2006}), \eprint{gr-qc/0510067}.

\bibitem[{\citenamefont{Chatziioannou et~al.}(2012)\citenamefont{Chatziioannou,
  Yunes, and Cornish}}]{Chatziioannou:2012rf}
\bibinfo{author}{\bibfnamefont{K.}~\bibnamefont{Chatziioannou}},
  \bibinfo{author}{\bibfnamefont{N.}~\bibnamefont{Yunes}}, \bibnamefont{and}
  \bibinfo{author}{\bibfnamefont{N.}~\bibnamefont{Cornish}},
  \bibinfo{journal}{Phys. Rev. D} \textbf{\bibinfo{volume}{86}},
  \bibinfo{pages}{022004} (\bibinfo{year}{2012}), \bibinfo{note}{[Erratum:
  Phys.Rev.D 95, 129901 (2017)]}, \eprint{1204.2585}.

\end{thebibliography}

\end{document}